\documentclass[trackchanges, twocolumn]{aastex7}

\hypersetup{pdfauthor={Name}}
\usepackage{caption}

\begin{document}
\title{The ALMA-QUARKS Survey: Evidence of an Explosive Molecular Outflow in IRAS 15520--5234}
\correspondingauthor{Ariful Hoque}

\author[0009-0003-6633-525X]{Ariful Hoque}
\affiliation{S. N. Bose National Centre for Basic Sciences, Block-JD, Sector-III, Salt Lake City, Kolkata 700106, India}
\email[show]{arifulh882@gmail.com} 

\author[0000-0003-0295-6586]{Tapas Baug}
\affiliation{S. N. Bose National Centre for Basic Sciences, Block-JD, Sector-III, Salt Lake City, Kolkata 700106, India}
\email[]{tapas.polo@gmail.com} 

\author[0000-0003-2630-3774]{Estrella Guzman}
\affiliation{Instituto Argentino de Radioastronomía (CCT-La Plata, CONICET, UNLP, CICPBA), C.C. No. 5, 1894, Villa Elisa, Buenos Aires, Argentina}
\email[]{estreguzman@gmail.com} 

\author[0000-0001-5811-0454]{Manuel Fernandez Lopez}
\affiliation{Instituto Argentino de Radioastronomía (CCT-La Plata, CONICET, UNLP, CICPBA), C.C. No. 5, 1894, Villa Elisa, Buenos Aires, Argentina}
\email[]{manferna@gmail.com} 

\author[0000-0002-5286-2564]{Tie Liu}
\affiliation{Shanghai Astronomical Observatory, Chinese Academy of Sciences, 80 Nandan Road, Shanghai 200030, China}
\affiliation{Key Laboratory for Research in Galaxies and Cosmology, Shanghai Astronomical Observatory, Chinese Academy of Sciences, 80 Nandan Road, Shanghai 200030, China}
\email[]{liutie@shao.ac.cn}

\author[0000-0003-1649-7958]{Guido Garay}
\affiliation{Departamento de Astronomía, Universidad de Chile, Las Condes, Santiago 7550000, Chile}
\affiliation{Chinese Academy of Sciences South America Center for Astronomy, National Astronomical Observatories, CAS, Beijing 100101, China}
\email[]{guido@das.uchile.cl}

\author[0000-0002-6622-8396]{Paul F. Goldsmith}
\affiliation{Jet Propulsion Laboratory, California Institute of Technology, 4800 Oak Grove Drive, Pasadena, CA 91109, USA}
\email[]{Paul.F.Goldsmith@jpl.nasa.gov}

\author[0000-0001-5950-1932]{Fengwei Xu}
\affiliation{Max Planck Institute for Astronomy, Königstuhl 17, 69117 Heidelberg, Germany}
\email[]{fengwei@mpia.de, fengweilookuper@gmail.com}

\author[0000-0002-4154-4309]{Xindi Tang}
\affiliation{XingJiang Astronomical Observatory, Chinese Academy of Sciences(CAS), Urumqi 830011, PR China}
\email[]{tangxindi@xao.ac.cn}

\author[0000-0002-7125-7685]{Patricio Sanhueza}
\affiliation{Department of Astronomy, School of Science, The University of Tokyo, 7-3-1 Hongo, Bunkyo, Tokyo 113-0033, Japan}
\email[]{patosanhueza@gmail.com}

\author[0000-0001-6725-0483]{Lokesh K. Dewangan}
\affiliation{Physical Research Laboratory, Navrangpura, Ahmedabad 380009, India}
\email[]{loku007@gmail.com}

\author[0000-0002-8614-0025]{Shivani Gupta}
\affiliation{Indian Institute of Astrophysics, II Block, Koramangala, Bengaluru 560034, India}
\affiliation{Pondicherry University, R.V. Nagar, Kalapet, 605014, Puducherry, India}
\email[]{shivani.gupta@iiap.res.in}

\author[0000-0002-8697-9808]{Sami Dib}
\affiliation{Max Planck Institute for Astronomy, K\"{o}nigstuhl 17, 69117, Heidelberg, Germany}
\email[]{sami.dib@gmail.com}

\author[0000-0003-2343-7937]{Luis A. Zapata}
\affiliation{Instituto de Radioastronom\'ia y Astrof\'isica, Universidad Nacional Autónoma de México, 58090, Morelia, Michoacán, México}
\email[]{l.zapata@crya.unam.mx}

\author[0000-0001-7866-2686]{Jihye Hwang}
\affiliation{Institute for Advanced Study, Kyushu University, Japan}
\affiliation{Department of Earth and Planetary Sciences, Faculty of Science, Kyushu University, Nishi-ku, Fukuoka 819-0395, Japan}
\email[]{hwang.jihye.514@m.kyushu-u.ac.jp}

\author[0000-0001-8812-8460]{N. K. Bhadari}
\affiliation{Kavli Institute for Astronomy and Astrophysics, Peking University, 5 Yiheyuan Road, Haidian District, Beijing 100871, China}
\email[]{naval1996kishor@gmail.com}

\author[0000-0001-8135-6612]{John Bally}
\affiliation{CASA, University of Colorado, 389-UCB, Boulder, CO 80309, USA}
\email[]{john.bally@colorado.edu}

\author[0000-0002-3658-0516]{Swagat Ranjan Das}
\affiliation{ Departamento de Astronomía, Universidad de Chile, Las Condes, Santiago 7550000, Chile}
\email[]{swagat@das.uchile.cl}

\author[0000-0003-4546-2623]{Aiyuan Yang}
\affiliation{National Astronomical Observatories, Chinese Academy of Sciences, Beijing 100101, People's Republic of China}
\affiliation{Key Laboratory of Radio Astronomy and Technology, Chinese Academy of Sciences, A20 Datun Road, Chaoyang District, Beijing, 100101, People's Republic of China}
\email[]{yangay@bao.ac.cn}

\author[0000-0003-1602-6849]{Prasanta Gorai}
\affiliation{Rosseland Centre for Solar Physics, University of Oslo, PO Box 1029 Blindern, 0315 Oslo, Norway}
\affiliation{Institute of Theoretical Astrophysics, University of Oslo, PO Box 1029 Blindern, 0315 Oslo, Norway}
\email[]{prasanta.astro@gmail.com}

\author[0000-0002-7367-9355]{Arup Kumar Maity}
\affiliation{Physical Research Laboratory, Navrangpura, Ahmedabad 380009, India}
\email[]{aruokumarmaity123@gmail.com}

\author[0000-0002-9875-7436]{James O. Chibueze}
\affiliation{UNISA Centre for Astrophysics and Space Sciences (UCASS), College of Science, Engineering and Technology, University of South Africa, Cnr Christian de Wet Rd and Pioneer Avenue, Florida Park, 1709, Roodepoort, South Africa}
\affiliation{Centre for Space Research, North-West University, Potchefstroom 2520, South Africa}
\affiliation{Department of Physics and Astronomy, Faculty of Physical Sciences, University of Nigeria, Carver Building, 1 University Road, Nsukka 410001, Nigeria}
\email[]{james.chibueze@gmail.com}

\author[0000-0002-8586-6721]{Pablo Garc\'ia}
\affiliation{Chinese Academy of Sciences South America Center for Astronomy, National Astronomical Observatories, CAS, Beijing 100101, China}
\affiliation{Instituto de Astronomía, Universidad Católica del Norte, Av. Angamos 0610, Antofagasta, Chile}
\email[]{pablo.garcia@nao.cas.cn}

\author[0000-0002-9574-8454]{Leonardo Bronfman}
\affiliation{Departamento de Astronomía, Universidad de Chile, Las Condes, Santiago 7550000, Chile}
\email[]{leo@das.uchile.cl}

\author[0000-0001-8315-4248]{Xunchuan Liu}
\affiliation{Shanghai Astronomical Observatory, Chinese Academy of Sciences, 80 Nandan Road, Shanghai 200030, China}
\email[]{liuxunchuan@shao.ac.cn}

\author[0000-0001-7598-9026]{L. Viktor Tóth}
\affiliation{University of Debrecen, Institute of Physics, H-4032 Debrecen, Bem tér 1.}
\email[]{toth.laszlo.viktor@ttk.elte.hu}

\author[0009-0002-5015-9979]{Shehu Muhammad Usman}
\affiliation{S. N. Bose National Centre for Basic Sciences, Block-JD, Sector-III, Salt Lake City, Kolkata 700106, India}
\email[]{mu.shehu@kasu.edu.ng}

\author[0000-0003-2412-7092]{Kee-Tae Kim}
\affiliation{Korea Astronomy and Space Science Institute, 776 Daedeokdae-ro, Yuseong-gu, Daejon 34055, Republic of Korea}
\affiliation{University of Science and Technology, Korea (UST), 217 Gajeong-ro, Yuseong-gu, Daejeon, 34113,  Republic of Korea}
\email[]{ktkim@kasi.re.kr}

\begin{abstract}
We present a study of the massive protocluster IRAS 15520--5234, which displays evidence of an explosive molecular outflow that unleashed a kinetic energy of at least 10$^{48}$ erg.
The protocluster contains 16 dense cores detected in the ALMA band 6 continuum emission maps, having masses in the range from 0.2 to 11.0 M$_\odot$. Our analysis of CO $(2-1)$ emission reveals 28 well collimated outflow fingers, the majority of which follow a Hubble-Lemaître velocity law. The outflow fingers show no preferred orientation in the plane of sky and emerge from a common center of origin. We estimate the total mass, momentum, and kinetic energy of the outflow fingers and find that the values are at least one order of magnitude higher than the typical bipolar outflows associated with massive protostars. The morphology and kinematics of the outflow fingers suggest that the outflow associated with IRAS 15520--5234  is explosive in nature. We calculate the dynamical age of the explosive event to be approximately 6550 years. Additionally, we estimate the frequency of such explosive outﬂows in the Galaxy, which is one event every 83 years. Finally, we speculate that the rearrangement of masses within the massive protocluster and the dynamical interaction among the massive cores may result in the formation of such an energetic event.
\end{abstract}
%\keywords{Interstellar molecules (849) --- Star forming regions (1565) --- }
\section{Introduction}\label{sec:introduction}
Massive protoclusters are the birthplaces of most of the massive stars and play a crucial role in their formation and evolution within molecular clouds. These massive protoclusters are often associated with highly energetic outflows that act as a primary mechanism for the release of gravitational potential energy. During the initial phases of protostellar evolution, two types of molecular outflows are observed - the classical bipolar outflows and explosive outflows. Bipolar outflows are generated through active accretion onto the central protostar via a flattened disk. These types of outflows are observed in both low-mass and high-mass star-forming regions \citep[][and references therein]{bachiller1996,lopez2009, arce2010, maud2015, baug2020, guerra2023, towner2024, hoque2025}. Explosive outflows are a distinct and highly energetic category of molecular outflows \citep[][]{Zapata2009, Zapata2013, zapata2019, guzman2022, bally2022}. They are characterized by multiple narrow filaments emerging out of a common center and follow a Hubble-Lemaître velocity law, i.e., linearly increasing velocity of the outflowing gas with distance from the common center.   
While bipolar outflows are studied extensively in both low- and high-mass star formation, the study of explosive outflows is limited due to their rarity and shorter timescales.  To date, only seven explosive outflow events have been reported in our Galaxy, and all of them are associated with massive star-forming regions \citep[][]{Zapata2009, Zapata2013, zapata2019, guzman2022, bally2022, zapata2023, issac2025}.

Multiple theoretical and analytical models have been proposed and tested to explain the origin of explosive outflows \citep[][]{Bally2011, rivera2021, raga2021, rivera2025}. For example, \citet[][]{Bally2011} proposed that the rearrangement of non-hierarchical systems and their dynamical interactions can lead to the formation of explosive outflows.  
\citet[][]{rivera2021}  performed an N-body simulation that accounts for the close encounter of a massive runaway star (of mass 10 M$_{\odot}$) with a cluster of massive particles and found that when the mass of the cluster is less than or up to a few times of the stellar mass, the collision between them will produce an explosive outflow.

The massive protocluster IRAS 15520--5234 (hereafter I15520) is located at a distance of 2.56 kpc and have a local standard of rest velocity, V$_{\text{lsr}}$ of -41.8 km s$^{-1}$ \citep[][]{xunchuan2024}. The associated massive clump has mass and luminosity of 10$^{3.2}$ M$_{\odot}$ and 10$^{5.1}$ L$_{\odot}$, respectively \citep[][]{Urquhart2017,liu2020}. The protocluster is also associated with evolved ultracompact (UC) H{\sc ii} regions \citep{ellingsen2005}. Molecular outflows have been identified with the Atacama Large Millimeter/submillimeter Array (ALMA) band 7 at 0.87 mm by \citet[][]{baug2020} in the region. The authors have identified a total of 7 outflow lobes (5 blue-shifted and 2 red-shifted) associated with 0.87 mm continuum cores. In another study by \citet[][]{Xu2024}, multiple prestellar and protostellar cores were identified in the region with ALMA band 7. Our recent ALMA-ATOMS survey of HC$_3$N outflows has revealed the presence of multiple outflow lobes (4 blue-shifted and 3 red-shifted) emerging out of a single core (at an angular resolution of $\sim1\farcs7$), hinting to an explosive outflow event in I15520 \citep[][]{hoque2025}. 

In this study, we utilized high-resolution ALMA band 6 data from the QUARKS survey and detected 28 outflow fingers associated with I15520. We confirm the explosive nature of the outflow fingers and investigate the possible driving mechanism of the explosive event.
The paper is organized in the following manner. In Section~\ref{sec:data}, we introduce the ALMA data and other archival data used in this study. Section~\ref{sec:results} presents the identification of continuum cores, outflow fingers, the study of outflow morphology and kinematics, and the radio emissions associated with the explosive outflow. In section~\ref{sec:discussion}, we present a discussion about the explosive nature of the outflow associated with I15520, the frequency of occurrence of such energetic events in our Galaxy, and the possible driving mechanism of the explosive event. Finally, a summary of this work is presented in section~\ref{sec:summary}.

\section{Observations and Data} \label{sec:data}
\subsection{ALMA Observations}
In this study, we utilize ALMA band 6 data from the QUARKS \citep[Querying Underlying Mechanisms of Massive Star Formation with ALMA-Resolved Gas Kinematics and Structures;][]{xunchuan2024} survey. The QUARKS survey includes 139 massive star-forming regions, observed using both the ALMA 12-m compact array and ACA 7-m array in ALMA band 6. More comprehensive details of the observations and data reductions can be found in \citet{xunchuan2024}. In our study, we use the 12-m array + ACA 7-m array combined data, which has a beam size of $\sim0\farcs36 \times 0\farcs33$
and a rms noise level of 0.2 mJy beam$^{-1}$. 
The details of the molecular line data from the QUARKS survey used in this study are mentioned in Table~\ref{tab:data}.%The molecular line data from the QUARKS survey used in this study have a spectral resolution of $\sim$1.3 km s$^{-1}$.

\begin{deluxetable}{ccccccc}
{
%\bfseries
\tabletypesize{\scriptsize}
\tablecaption{\text{Summary of the Spectral Lines from ALMA-QUARKS Survey Analyzed in this Study}  \label{tab:data}}
\tablehead{
\colhead{Molecule} & \colhead{Transition} & \colhead{Rest Frequency}  & \colhead{Synthesized Beam Size} & \colhead{Spectral Resolution} & rms per 0.488 MHz Channel\\ %& \colhead{Remarks}\\
\colhead{} & \colhead{} & \colhead{(GHz)} & \colhead{($''\times''$)} & \colhead{(km s$^{-1}$)} &  \colhead{(mJy beam$^{-1}$)} %& \colhead{}
}
\startdata
SiO & $5-4$ & 217.104919 & 0.41$\times$0.36 & 1.35 & 6.2\\
CH$_{3}$OH & $4(2,2)-3(1,2)$ & 218.440063 & 0.41$\times$0.36 & 1.34 & 5.8\\
SO & $6(5)-5(4)$ & 219.949442 & 0.42$\times$0.40 & 1.33 & 6.7\\
$^{13}$CO & $2-1$ & 220.398684 & 0.42$\times$0.39 & 1.32 & 7.8 \\
$^{12}$CO & $2-1$ & 230.538000 & 0.40$\times$0.37 & 1.27 & 8.2 \\
%$^{13}$CS & $5-4$ & 231.220685 & 0.40$\times$0.37 & 1.27 & 7.2\\
H$_{30\alpha}$ & & 231.900928 & 0.40$\times$0.37 & 1.26 & 6.0 \\
\enddata
}
\end{deluxetable}

\subsection{Archival Data}
The {\it Spitzer} Space Telescope Infrared Array Camera (IRAC) Ch1 (3.6 $\mu$m) and Ch2 (4.5 $\mu$m) archival images were obtained from the Galactic Legacy Infrared Mid-Plane Survey Extraordinaire survey \citep[GLIMPSE;][]{benjamin2003}. The {\it Spitzer}-IRAC images have a spatial resolution of $\sim2\farcs0$ and a sensitivity of $\sim0.2-0.3$ mJy beam$^{-1}$.

We utilize the 5.5 GHz radio continuum image from the Co-Ordinated Radio `N' Infrared Survey for High-mass star formation \citep[CORNISH-SOUTH;][]{irabor2023} survey to identify the ionized emissions in the region. The CORNISH-SOUTH image has an angular resolution of $\sim2\farcs5$ and a rms noise of $\sim0.11$ mJy beam$^{-1}$.
\section{Results} \label{sec:results}

\subsection{1.3 mm Continuum Emission}\label{sec:core identification}
\begin{figure*}[ht!]
\centering
    \includegraphics[width=0.8\textwidth]{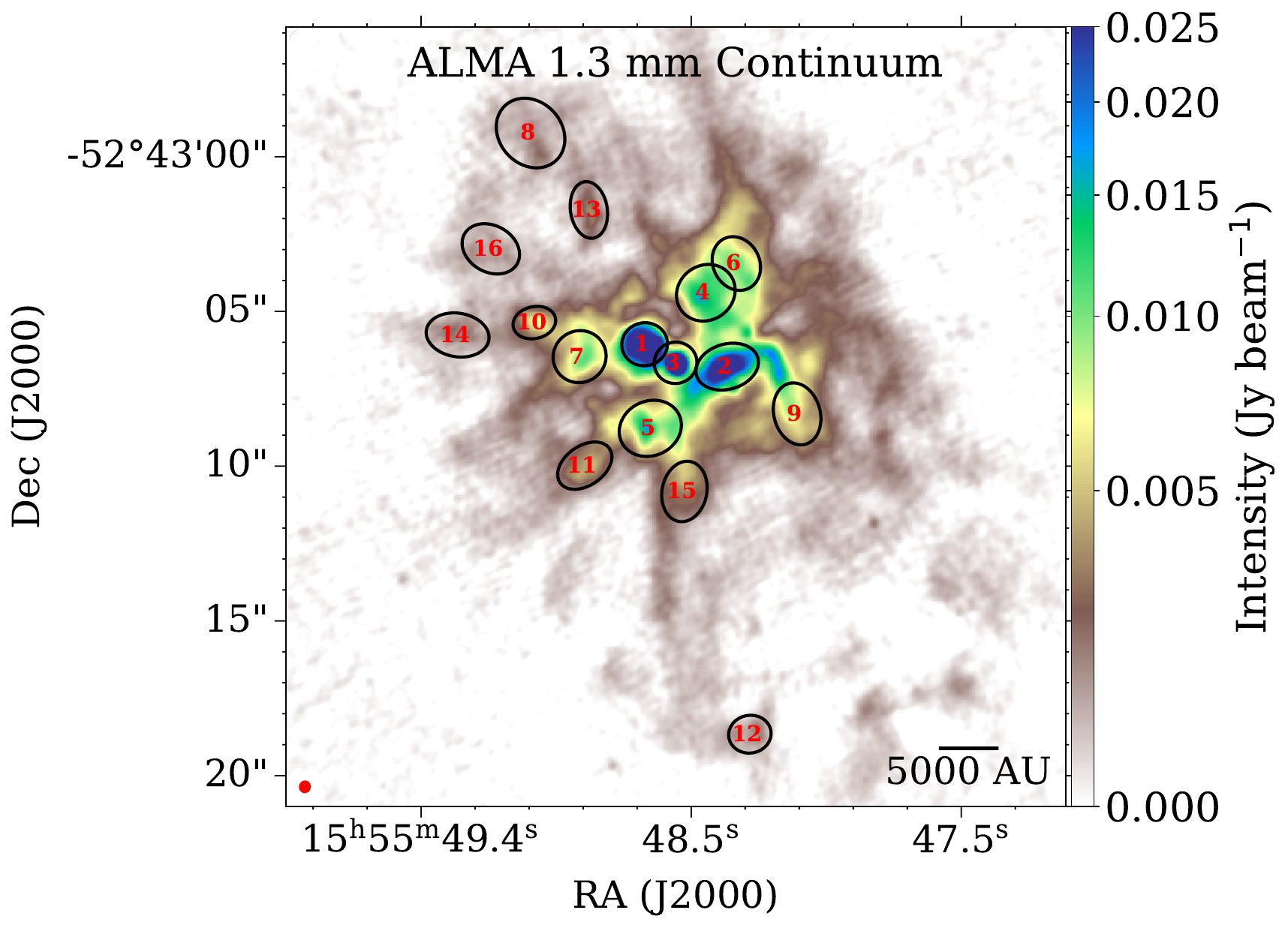}
\caption{ALMA 1.3 mm continuum emission in the region I15520. The black ellipses represent the cores identified by \citet[][]{yang2025} using the {\it getsf} algorithm. The cores are labeled numerically (in red) in the order they were identified in {\it getsf}. The beam is shown in the bottom-left corner, and the scale bar is shown in the bottom-right corner.}
\label{fig-continuum}
\end{figure*}

In Figure~\ref{fig-continuum}, we present the 1.3 mm continuum emission maps for I15520, overlaid with the compact cores identified previously in the ALMA-QUARKS survey \citep[QUARKS-III;][]{yang2025}. In QUARKS-III, the \textit{getsf} source extraction algorithm \citep{Menshchikov2021} was adopted to extract the cores from the high-resolution ALMA 1.3 mm continuum map. Using molecular line data, the cores were further classified into different evolutionary stages, starting from starless cores to UCH{\sc ii} regions. A total of 16 compact cores were identified in the region. Among them, 10 were classified as hot molecular cores (HMCs) and the remaining 6 were classified as UCH{\sc ii} regions. The authors estimated the masses of the cores assuming a dust temperature of 100 K for the evolved cores \citep[see,][for details]{yang2025}. Note that, the 1.3 mm continuum map might be affected by free-free emissions and this could lead to the overestimation of the masses of the continuum cores. Table~\ref{tab:core} lists the details of the continuum cores associated with I15520. 

\begin{deluxetable*}{cccccccccc}
\tabletypesize{\scriptsize}
\tablecaption{Parameters of the 1.3 mm Continuum Cores \label{tab:core}}
\tablehead{
\colhead{Core No.} & \colhead{RA (J2000)} & \colhead{DEC (J2000)}  & \colhead{FWHM$_{maj}$} & \colhead{FWHM$_{min}$} & \colhead{PA} & \colhead{$\text{F}_{int}^{\dagger}$} & \colhead{Core Mass$^{\dagger}$} & \colhead{Evolutionary Stage}\\
{} & (hh:mm:ss) & (dd:mm:ss) & ($''$) & ($''$) &($^{\circ}$)& (mJy) &  (M$_{\odot}$) & {}
}
\startdata
1 & 15:55:48.65 & -52:43:06.10 & 1.482 & 1.388 & 100.3 & 362.2(3.9) & 8.16(3.29) & UCH{\sc ii}\\
2 & 15:55:48.36 & -52:43:06.82 & 2.061 & 1.459 & 106.3 & 484.3(7.4) & 10.95(4.43) & UCH{\sc ii}\\
3 & 15:55:48.54 & -52:43:06.70 & 1.387 & 1.338 & 14.2 & 220.4(3.9) & 4.95(2.05) & UCH{\sc ii}\\
4 & 15:55:48.44 & -52:43:04.44 & 1.953 & 1.769 & 34.7 & 153.8(5.8) & 3.48(1.46) & HMC\\
5 & 15:55:48.63 & -52:43:08.82 & 2.055 & 1.766 & 115.0 & 98.6(5.5) & 2.24(0.92) & UCH{\sc ii}\\
6 & 15:55:48.33 & -52:43:03.49 & 1.791 & 1.500 & 26.8 & 60.3(3.9) & 1.37(0.56) & HMC\\
7 & 15:55:48.89 & -52:43:06.50 & 1.724 & 1.670 & 118.1 & 81.6(4.0) & 1.84(0.78) & HMC\\
8 & 15:55:49.06 & -52:42:59.27 & 2.422 & 2.026 & 42.6 & 36.2(2.2) & 0.82(0.35) & HMC\\
9 & 15:55:48.11 & -52:43:08.35 & 2.037 & 1.495 & 15.3 & 39.5(4.3) & 0.89(0.38) & UCH{\sc ii}\\
10 & 15:55:49.05 & -52:43:05.40 & 1.397 & 1.034 & 102.7 & 20.9(3.0) & 0.47(0.20) & HMC\\
11 & 15:55:48.87 & -52:43:10.02 & 1.944 & 1.277 & 125.0 & 21.4(2.6) & 0.48(0.21) & HMC\\
12 & 15:55:48.28 & -52:43:18.70 & 1.379 & 1.221 & 99.6 & 10.5(0.7) & 0.24(0.10) & HMC\\
13 & 15:55:48.85 & -52:43:01.76 & 1.847 & 1.189 & 8.8 & 11.1(1.3) & 0.25(0.11) & HMC\\
14 & 15:55:49.32 & -52:43:05.80 & 2.045 & 1.409 & 81.2 & 15.2(1.6) & 0.34(0.14) & HMC\\
15 & 15:55:48.51 & -52:43:10.86 & 1.972 & 1.445 & 168.6 & 9.5(2.0) & 0.21(0.10) & UCH{\sc ii}\\
16 & 15:55:49.20 & -52:43:03.02 & 1.957 & 1.502 & 60.4 & 9.2(1.7) & 0.21(0.09) & HMC
\enddata
\tablenotetext{\dagger}{The measurement uncertainty is shown in the parentheses.}
\tablerefs{\citet[][]{yang2025}}
\end{deluxetable*}

\subsection{Identification of Outflows and their Kinematics} \label{subsec:identification}

\begin{figure*}[ht!]
    \centering
        \includegraphics[width=0.49\textwidth]{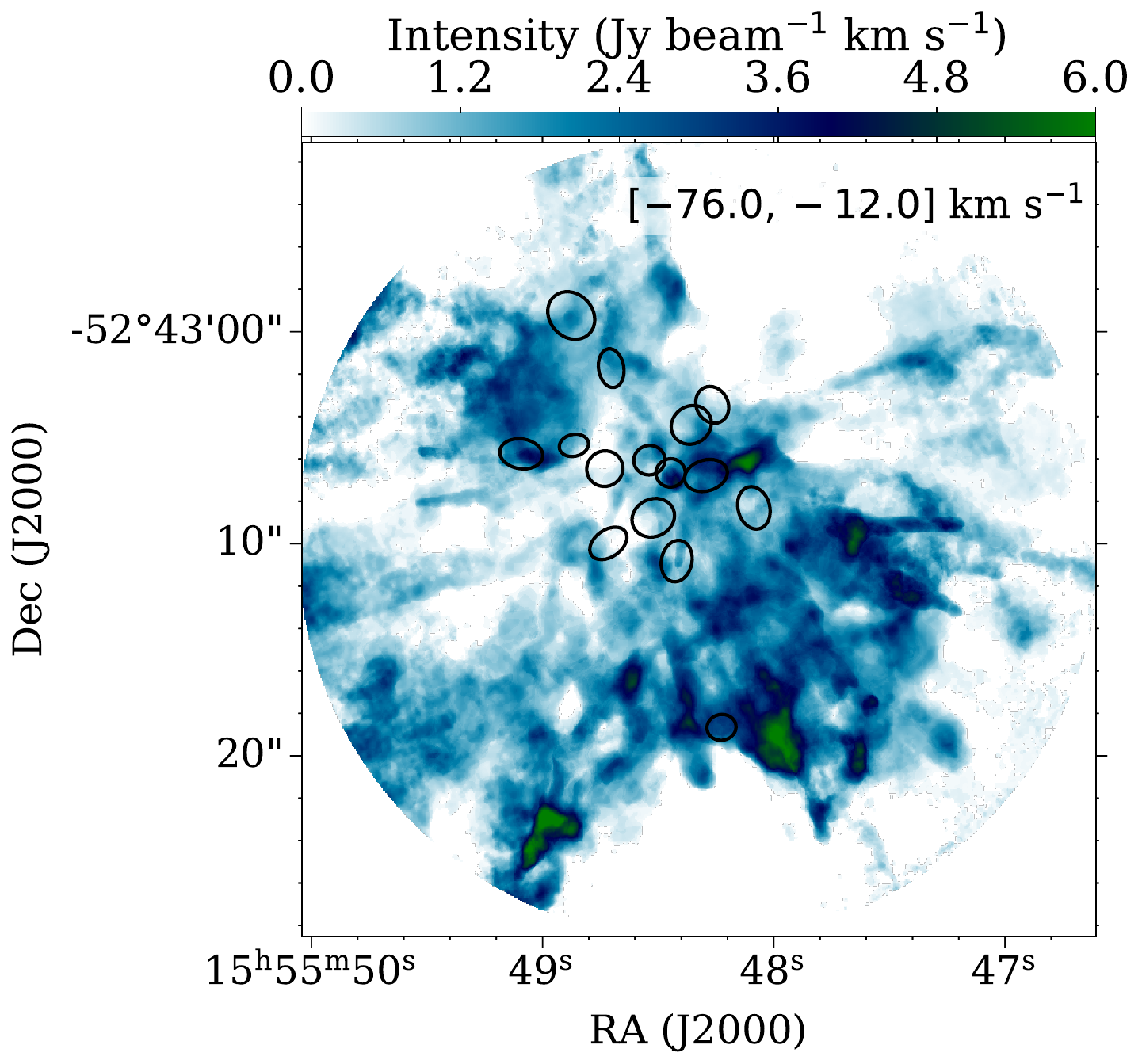}
        \includegraphics[width=0.5\textwidth]{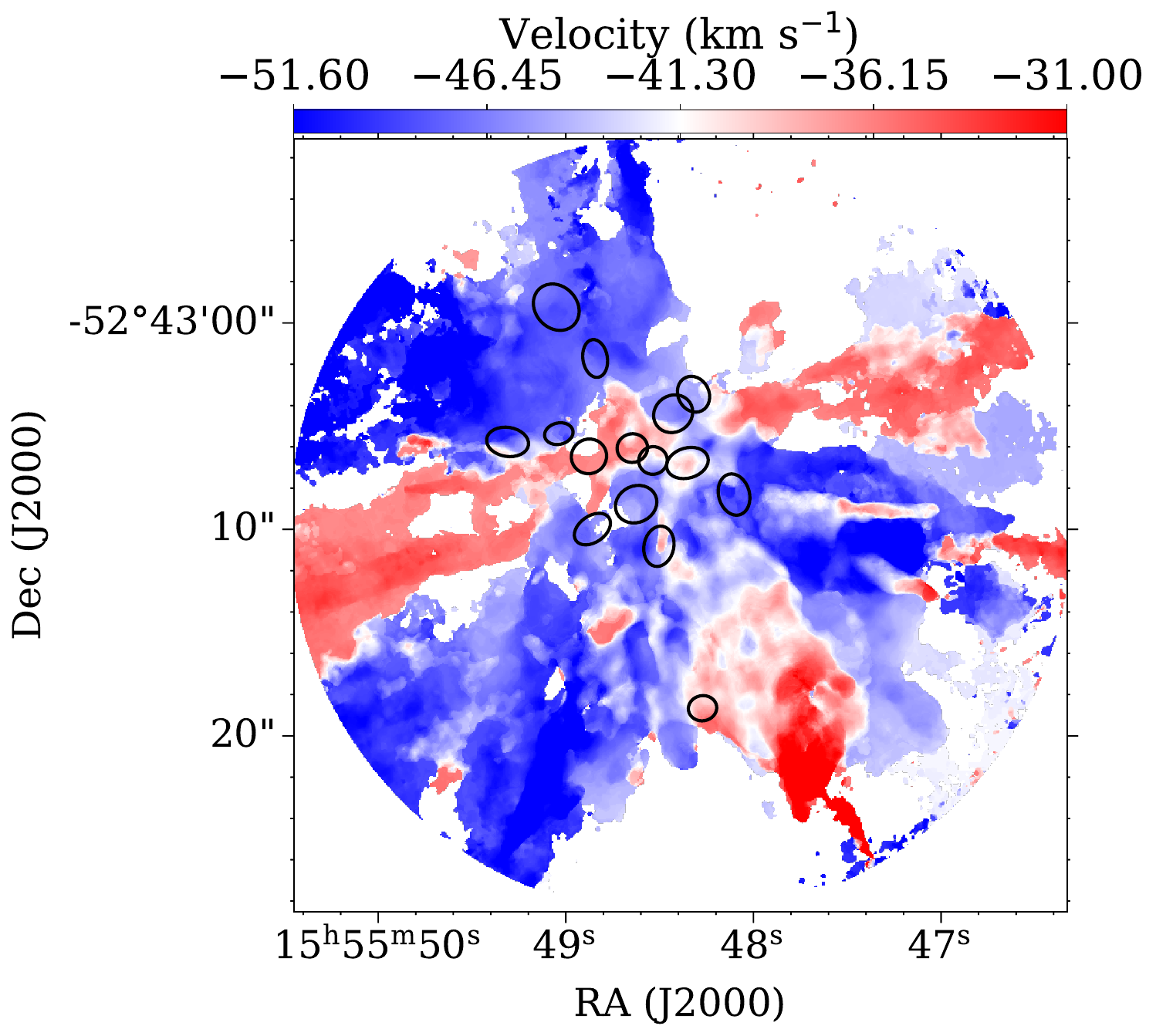}\\
        \includegraphics[width=0.45\textwidth]{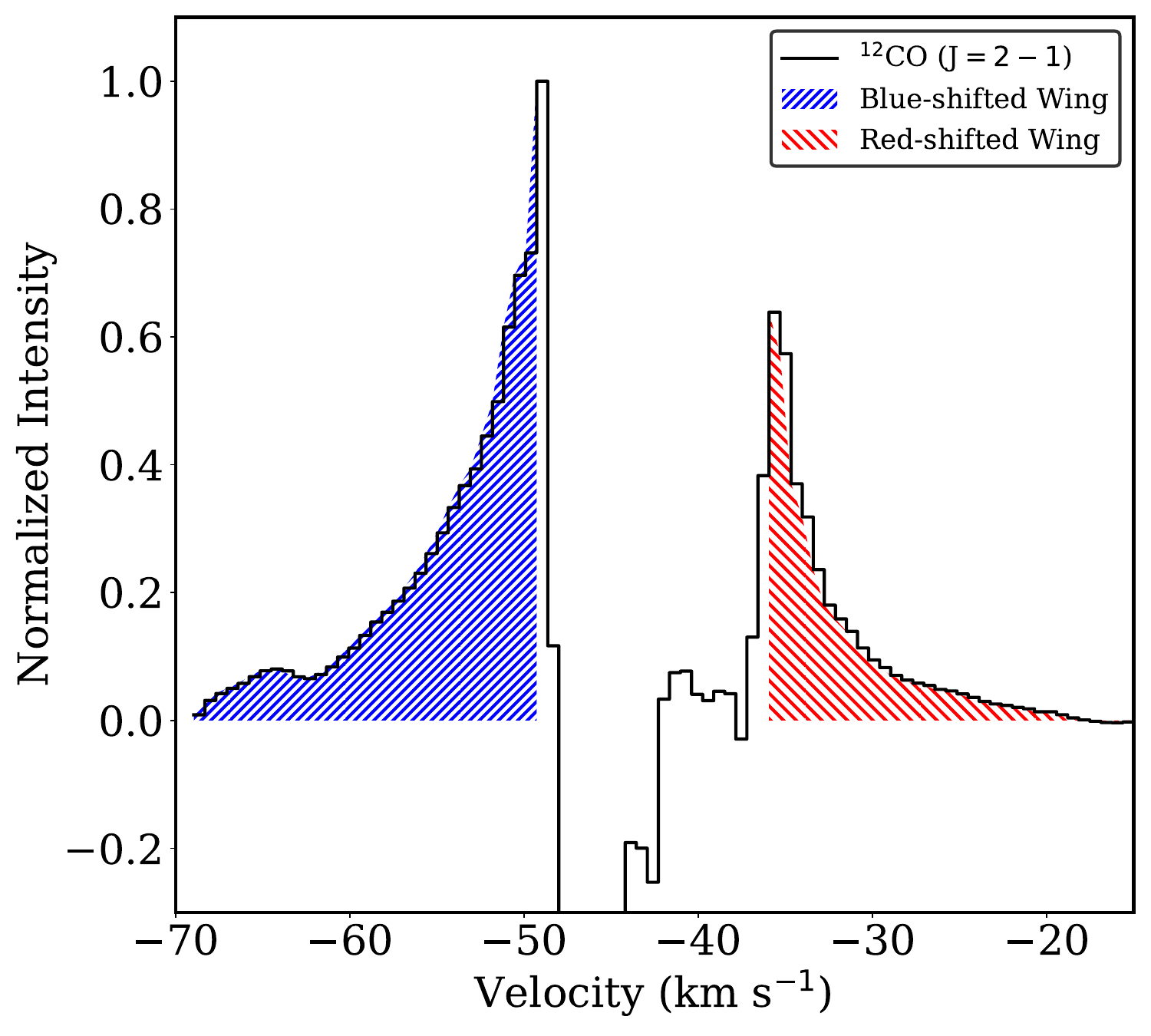}
    \caption{(Top) ALMA CO moment-0 (left) and moment-1 (right) maps of the region I15520. In each panel, the identified continuum cores are marked with black ellipses. The velocity range of integration is mentioned in the top right corner of the CO moment-0 map. (Bottom) Normalized CO spectrum integrated over the ALMA field of view (black). The blue and red hatched regions indicate blue- and red-shifted wing emissions, respectively.}
    \label{fig-spectra}
\end{figure*}

\begin{figure*}[h!]
\centering
    \includegraphics[width=0.49\textwidth]{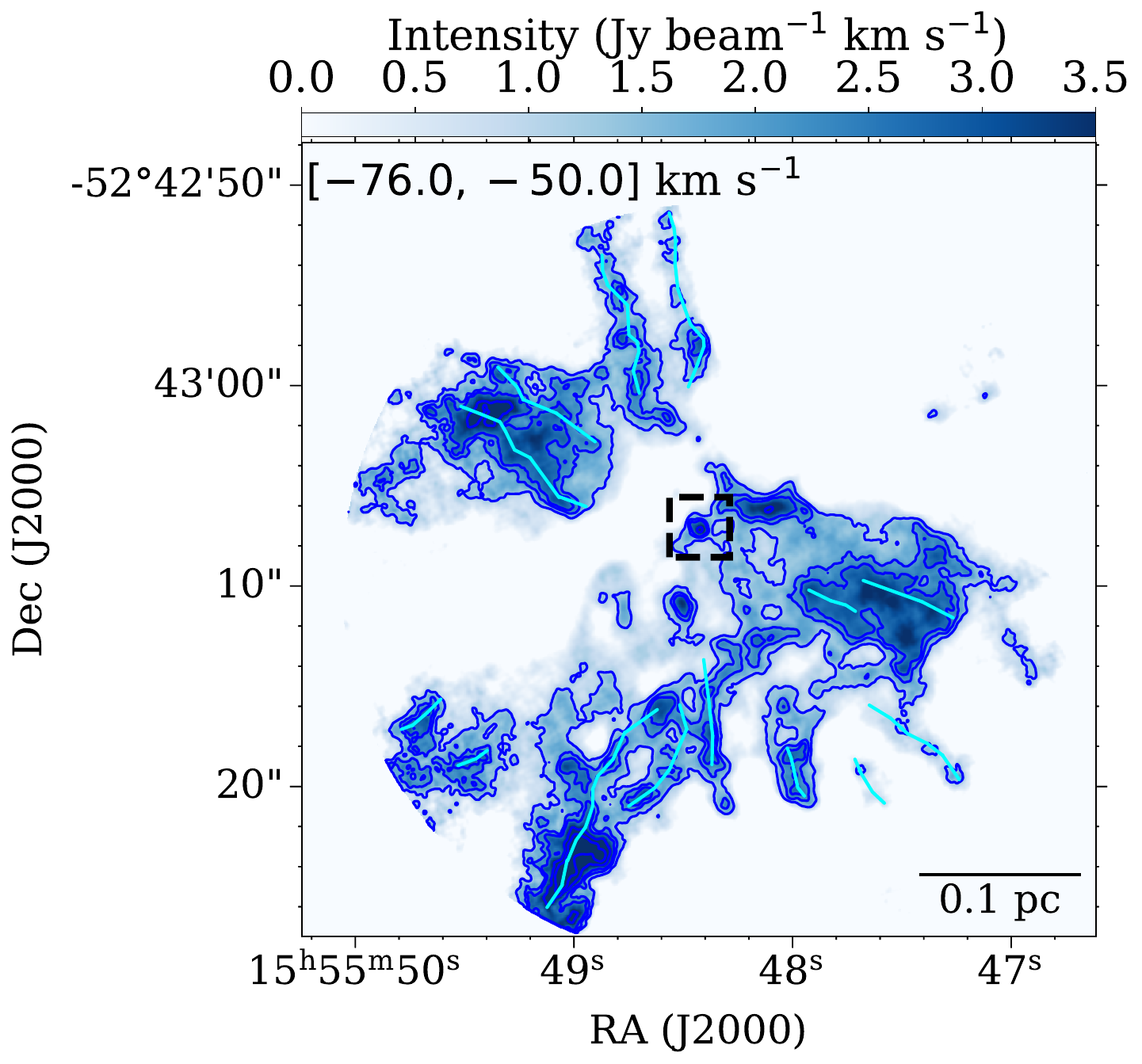}
    \includegraphics[width=0.49\textwidth]{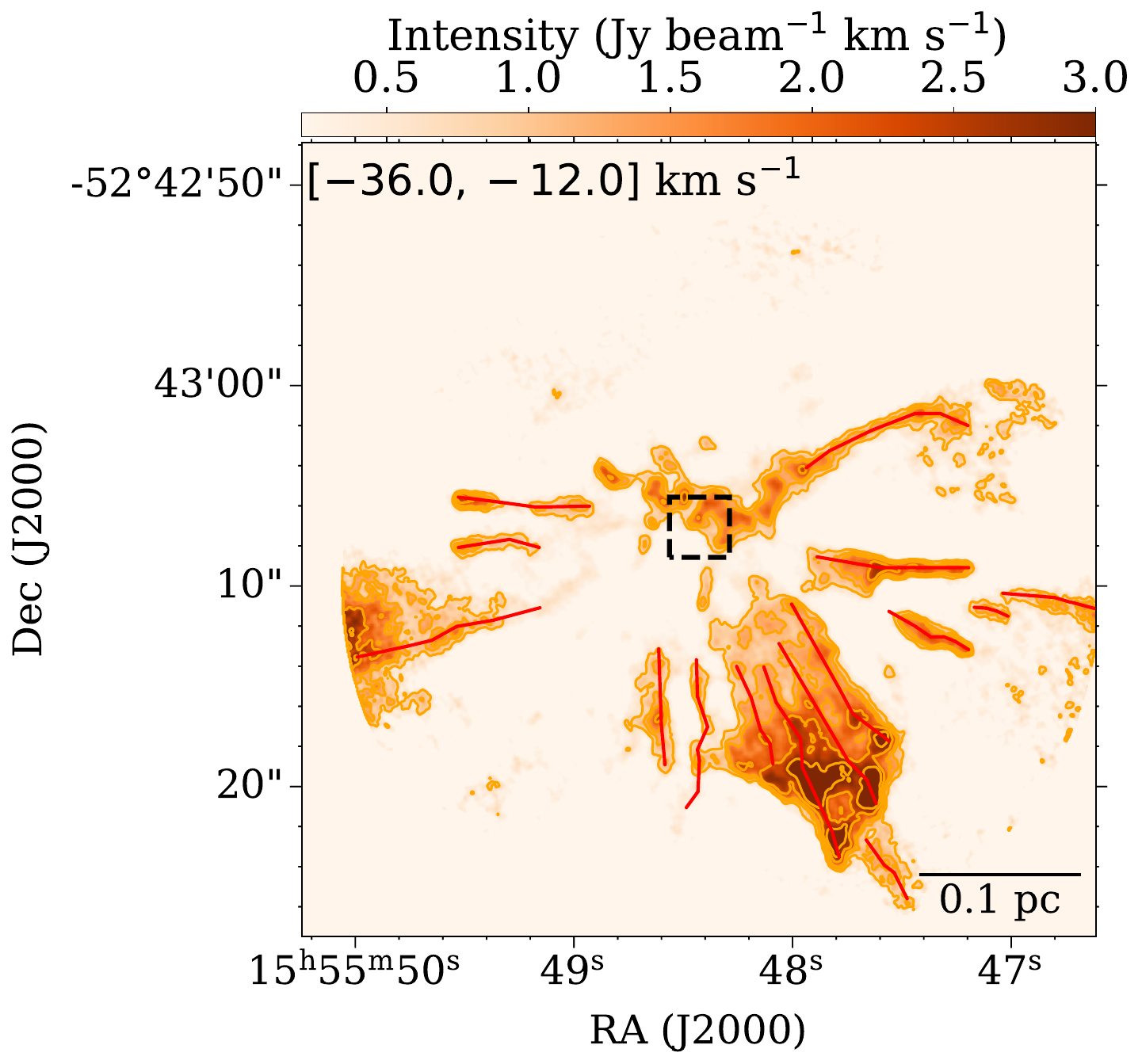}
\caption{The CO blue-shifted (left) and red-shifted (right) wing emission maps. The cyan and red lines represent the blue-shifted and red-shifted outflow fingers, respectively. The blue (in left) and red (in right) contours are drawn at the levels of [5, 8, 10, 15, 20, 50]$\times\sigma$, where $\sigma= 0.15$ and 0.25 Jy beam$^{-1}$ km s$^{-1}$ for the blue- and red-shifted wing emission maps, respectively. The velocity ranges of the wing emission maps are given in the top left corner of each figure. The black dashed rectangle represents the region of interest from where the explosive event is possibly originating.}
\label{fig-wing emission map}
\end{figure*}

We utilize the $^{12}$CO (J$=2-1$) molecular line (hereafter CO) data from the ALMA band-6 observations for the identification and kinematic study of the outflow fingers. We generate the integrated intensity (i.e., moment-0), and the intensity-weighted velocity (i.e., moment-1) maps using the \texttt{spectral-cube} package from the \texttt{astropy} project \citep{astropy2013}. The top-left panel of Figure~\ref{fig-spectra} shows the CO moment-0 map, tracing the isotropic distribution of finger-like outflow structures, while the top-right panel presents the CO moment-1 map, tracing the velocity of these structures. The bottom panel of Figure~\ref{fig-spectra} shows the normalized spectra of the CO emissions marked with regions of blue- and red-shifted wing emissions in blue and red hatches.
The CO emissions show self-absorption toward the central velocity of the cloud. Therefore, to exclude the contamination of the central cloud and mitigate the effect of self-absorption, we exclude the velocity range $-50.0$ km s$^{-1}$ to $-36.0$ km s$^{-1}$ for the estimation of outflow parameters and kinematics. We generate the blue- and red-shifted outflow wing emission maps by integrating emission in the velocity ranges $[-76.0, -50.0]$ km s$^{-1}$ and $[-36.0, -12.0]$ km s$^{-1}$, respectively. We present the blue-shifted and red-shifted wing emission maps in Figure~\ref{fig-wing emission map}.

We further investigate the CO data by generating channel maps for the blue- and red-shifted outflow components in the velocity range $[-76.0, -50.0]$ km s$^{-1}$ and $[-36.0, -12.0]$ km s$^{-1}$, respectively, with each channel representing integrated intensity within 2 km s$^{-1}$ velocity width. The CO channel maps are presented in the Appendix~\ref{Appendix-ChannelMaps} (see Figure~\ref{Fig:Appendix- blueshifted channel maps} and ~\ref{Fig:Appendix- redshifted channel maps}). The outflow fingers in Figure~\ref{Fig:Appendix- blueshifted channel maps} and ~\ref{Fig:Appendix- redshifted channel maps} appear to be randomly oriented in the plane of sky.
Outflow fingers are identified by investigating consecutive velocity channels, starting from the outermost velocity channels in the blue- and red-shifted directions with respect to the V$_{\text{lsr}}$. We generate contour at emissions $\geq5\sigma$ (where $\sigma$ represents the rms of the background noise) in each velocity channel and trace the ``spines" along the emission peaks.
Finally, by aggregating the spines from all velocity channels, we map the trajectories of the outflow fingers. We find a total of 14 blue-shifted and 14 red-shifted narrow outflow fingers emerging from the central cloud. Although the outflow fingers are believed to originate from a common center, pinpointing the exact location is challenging, as some fingers show no emissions toward the cloud center. Therefore, we define a common center of origin at RA: 15h55m48.51s, Dec:-52d43m07.11s with a box of size $4\farcs5\times4\farcs5$ such that, when extended in backward directions, the outflow fingers approximately converge within this box.
In Figure~\ref{fig-wing emission map}, we mark the blue- and red-shifted outflow fingers with cyan and red lines and the location of the common center by a black dashed rectangle. 
We also search for the possibility of any bipolar outflows in the region by investigating the velocity opposite channels with respect to V$_{\text{lsr}}$ ($=-41.8$ km s$^{-1}$) in Figure~\ref{Fig:Appendix- blueshifted channel maps} and \ref{Fig:Appendix- redshifted channel maps}, and do not find any bipolar components associated with the outflow fingers.

\begin{figure*}[h!]
\centering
    \includegraphics[width=0.6\textwidth]{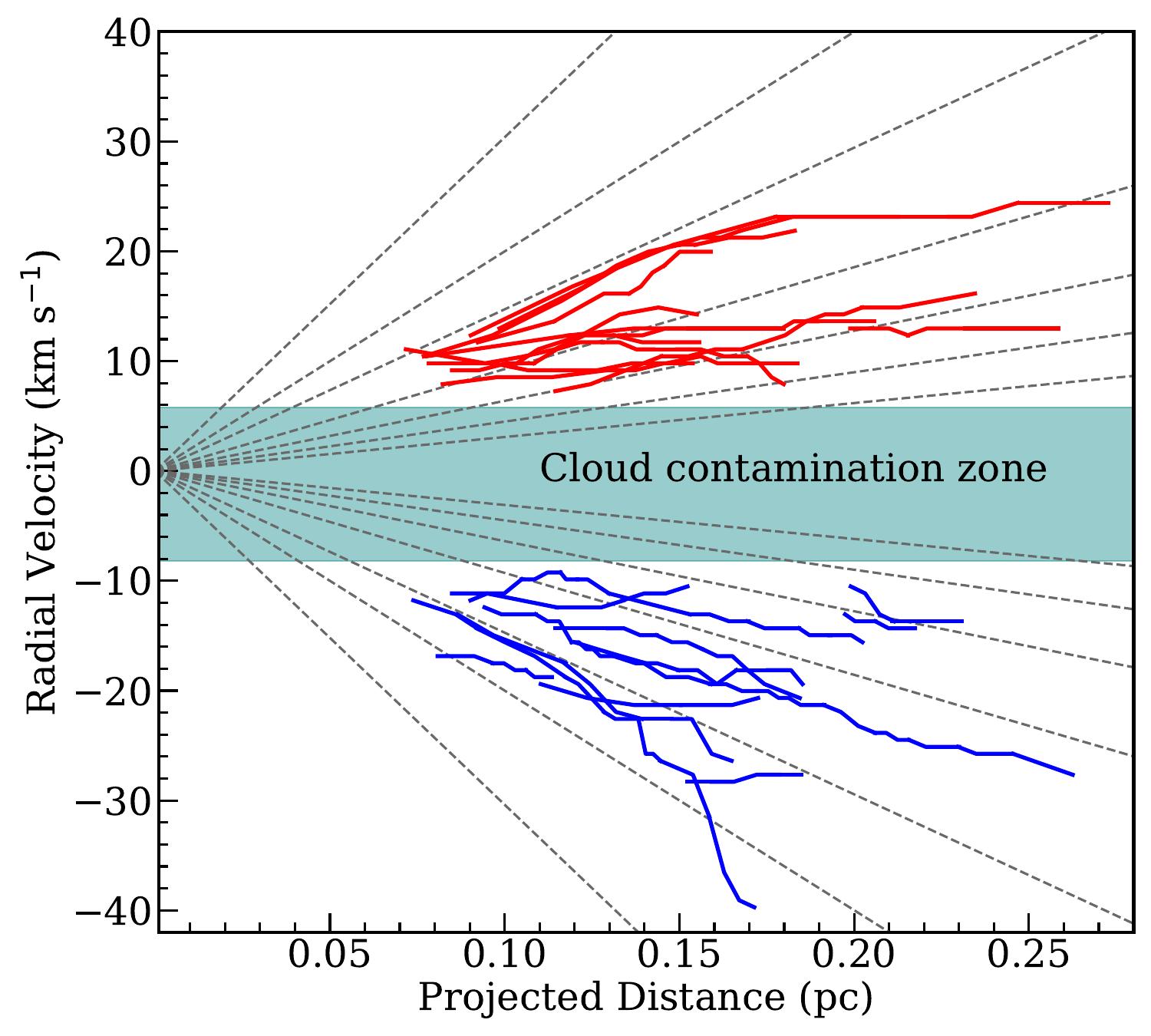}
\caption{The Position-Velocity (PV) diagram of the CO outflow fingers associated with I15520. The blue and red lines represent the PV of the blue-shifted and red-shifted fingers, respectively. The radial velocities are given with respect to the cloud velocity. The projected distances are the distances measured from the explosive center. The gray lines represent a linear relationship between the radial velocity and the projected distances.}
\label{fig-radial velocity}
\end{figure*}

We estimate the radial velocity of the outflow fingers from the farthest velocity channel having emissions $>5\sigma$ within the outflow fingers, where $\sigma$ is the rms of the background noise, estimated from emission-free regions.
Figure~\ref{fig-radial velocity} shows the radial velocity of the outflowing gas as a function of the sky-projected distance.
The blue and red lines in Figure~\ref{fig-radial velocity} represent the radial velocity of the outflowing gas of the blue and red-shifted fingers with distance from the common center. We find a linearly increasing trend of the radial velocity of the outflowing gas with distance (Hubble-Lemaître velocity law). The linearly increasing velocity of outflowing materials are believed to be originated due to ballistic or accelarated motion of the outflowing material, driven by momentum-conserving winds \citep[][]{shu1991, Li1996, Arce2001}. Since the outflow fingers are oriented isotropically in space, the effect of different inclination angle will be added in the Hubble-Lemaître velocity law of each outflow finger. As a result, the PV diagrams of the outflow fingers will appear to have different slopes as seen in Figure~\ref{fig-radial velocity}.

Note that such kinematics of the outflow fingers, i.e., linearly increasing velocity from a common center, having different slopes of the PV diagrams of different outflow fingers were earlier reported in explosive outflows \citep{Zapata2009, Zapata2013, zapata2019, guzman2022, bally2022, zapata2023, guzman2024} and are one of the most defining characteristics of this category of outflows. 

\subsection{Outflow Energetics}
 In this section, we estimate the physical parameters of the outflow, such as its mass, momentum, and energy. We first estimate the CO column density ($N_{\text{thin}}$) assuming the emission to be optically thin following the methodology devised in \citet[][]{garden1991} and \citet[][]{sanhueza2012}, and then corrected for optical depth of CO following \citet[][]{Goldsmith1999} and \citet[][]{Mangum2015}.
 
 With the assumptions for the CO emission to be optically thin and also in local thermodynamic equilibrium (LTE), we derive the CO column density following the equation \citep[adopted from][]{garden1991, sanhueza2012},  
\begin{equation}
N_{\text{thin}}= \frac{8\pi\nu^3}{c^3}\frac{Q_{\text{rot}}}{g_{\text{u}} A_{\text{ul}}} 
\frac{\exp(E_{\text{l}}/kT_{\text{ex}})}{1-\exp(-h\nu/kT_{\text{ex}} )}  \frac{\int T_{\text{B}} dv} {J(T_{\text{B}})-J(T_{\text{bg}})}~,
\end{equation}
%and 
with $J(T)$ being defined as,
\begin{equation}\label{equ:cden}
    J(T)= \frac{h\nu}{k} \frac{1}{e^{h\nu/kT}-1}~,
\end{equation}
where $\nu$ is the frequency of the transition, $c$ is the speed of light, $Q_{\text{rot}}$ is the partition function, $g_{\text{u}}$ is the statistical weight of the upper level, $A_{\text{ul}}$ is the Einstein coefficient for spontaneous emission, $E_{\text{l}}$ is the energy of the lower level, $T_{\text{ex}}$ is the excitation temperature, $T_{\text{B}}$ is the brightness temperature, and  $T_{\text{bg}}$=2.7 K is the cosmic microwave background temperature. In our analysis, we measure the integrated intensity over the velocity range of the outflows, which is approximately equal to $\int T_{\text{B}} dv$. In Equation~\ref{equ:cden}, we use a $T_{\text{ex}}$ of 50 K  for the outflow materials. In general, the temperature of the outflowing material lies within $\sim20-100$~K \citep[][]{hatchell1999, Shimajiri2009}.
In high-mass star-forming clumps, the gas kinetic temperature is found in the range from 30 K to 200 K with an average of 62$\pm$2 K \citep[][]{tang2018}. \citet[][]{izumi2024} also estimated the kinetic temperature of outflows for 12 massive infrared dark clouds (IRDCs) using formaldehyde (H$_2$CO) emissions, in the range 26 K to 300 K, having an average kinetic temperature of 75$\pm$50 K. 
 The values of $Q_{\text{rot}}$($=kT_{\text{ex}}/hB\,+\,1/3$=18.45), $g_{\text{u}}(=5)$, $A_{\text{ul}}(=6.911\times10^{-7}$ s$^{-1})$, $E_{\text{l}}(=5.53$ K) are obtained from the Cologne Database for Molecular Spectroscopy \citep[CDMS;][]{muller2001} and XCLASS database \citep{moller2017}.

We corrected the CO column density for not optically thin CO emission following the methodology devised in \citet[][]{Goldsmith1999} and \citet[][]{Mangum2015}.
\begin{equation}
N =N_{\text{thin}} \frac{\tau}{1-\exp(-\tau)},
\end{equation}
 where $\tau$ is the optical depth of CO.
 
 Assuming the two isotopologues $^{12}$CO $(2-1)$ and $^{13}$CO $(2-1)$ have same excitation temperature, we estimate $\tau$ \citep[following ][]{Qiu2009} from,
 \begin{equation}
     \frac{I(^{12}\text{CO})}{I(^{13}\text{CO})}=\frac{1-\exp(-\tau)}{1-\exp(-\tau/\chi)},
 \end{equation}
 where $\chi$ is the abundance ratio of $^{12}$CO $(2-1)$ and $^{13}$CO $(2-1)$ derived using the relation, $\chi= (7.5\pm1.9) \text{D}_{\text{GC}}+(7.6\pm12.9)$ \citep{Wilson1994}. D$_{\text{GC}}$ is the Galactocentric distance to the source which is 6.2 kpc for I15520 \citep[][]{liu2020}.
 
We estimate the outflow mass ($M_{\text{out}}$), momentum ($P_{\text{out}}$), energy ($E_{\text{out}}$) and dynamical timescale ($t_{\text{dyn}}$) using the following equations given in \citet[][]{lopez2009} and \citet[][]{Wang2011},
\begin{equation}
    M_{\text{out}}=\sum_i{M_{\text{i}}}=d^2 \left[\frac{\text{H}_2}{\text{CO}}\right] \mu_{\text{H}} \, \text{m}_{\text{H}} \, A \, \sum_i {N_{\text{i}}}~,
\end{equation}
\begin{equation}
    P_{\text{out}}=\sum_i{M_{\text{i}} v_{\text{i}}} ~,
\end{equation}
\begin{equation}
    E_{\text{out}}=\frac{1}{2} \sum_i{M_{\text{i}} v_{\text{i}}^2}~,
\end{equation}
\begin{equation}
    t_{\text{dyn}}= \frac{L_{\text{lobe}}}{v_{\text{lobe}}}~,
\end{equation}
where $M_{\text{i}}$ and $v_{\text{i}}$ are the mass and velocity of each channel within the velocity range of the outflow, $[\text{CO}/\text{H}_2]$ is the relative abundance of CO in comparison to H$_2$, $d$ is the distance to the source, $\mu_{\text{H}}$ is the mean molecular weight (adopted as 2.8), m$_{\text{H}}$ is the mass of of the Hydrogen atom, $A$ is the angular sky area subtended by a single pixel, $N_{\text{i}}$ is the column density for each velocity channel, $L_{\text{lobe}}$ is the extent of the outflow lobe from the explosion center and $v_{\text{lobe}}$ is the terminal velocity of the outflow lobe. The value of $[\text{CO}/\text{H}_2]$ is adopted as $10^{-4}$ \citep[][]{blake1987}.
The observed outflow velocity is the line-of-sight component of the actual velocity, and the length of the outflow lobe is the plane of sky projection of the actual length. Therefore, for a better estimate of the outflow momentum and energy, a correction for the inclination angle needs to be applied. In this study, we assumed a mean inclination angle, $\theta$ of 57.3$^{\circ}$ \citep[see][for detailed discussion]{dunham2014} of the outflow fingers with respect to the line of sight direction and corrected the values of the outflow momentum ($P_{\text{out}}$) and energy ($E_{\text{out}}$) by multiplying them with 1/$\cos\theta$ and 1/$\cos^2\theta$, respectively.
We estimate the mass, momentum, and energy of the explosive outflow by integrating the emissions of each channel over the velocity range of the explosive outflow. We included emissions $\geq5\sigma$ in each channel for the estimation of outflow parameters, where $\sigma$ is calculated from emission-free channels.
%We estimate the mass, momentum, and energy of the explosive outflow by summing the values from all outflow fingers.
We calculate the total mass, momentum, and energy of the outflow as 23.8 M$_{\odot}$, 3129.8 M$_{\odot}$ km s$^{-1}$, 4.1$\times10^{48}$ erg, respectively.

We estimate the dynamical timescale of the explosive event by assuming that outflow fingers are isotropic and have the same v$_{3D}$, and the difference in the observed extent and radial velocity is because of the different inclination angle of the outflow fingers. Therefore, the most likely estimate of the true extent of the explosive outflow would come from the longest projected outflow finger. Similarly, the most likely estimate of the true v$_{3D}$ would come from the fastest (in the radial velocity) projected outflow finger. We find the extend of the explosive outflow as 0.26 pc and the value of v$_{3D}$ as 38.5 km s$^{-1}$. The estimated dynamical age of the explosive outflow is 6550 years.

Note that the estimated outflow parameters involve several observational biases include the missing flux from interferometric observations, the distance uncertainty, the assumption of a constant outflow temperature, and a constant H$_2$ to CO abundance ratio. Overall, these factors may result in the underestimation or overestimation of the outflow parameters by a factor of two to three.
\subsection{Radio Emission and NIR Shocked Gas Associated with the Explosive Outflow}
\begin{figure*}[ht!]
\centering
    \includegraphics[width=0.49\textwidth]{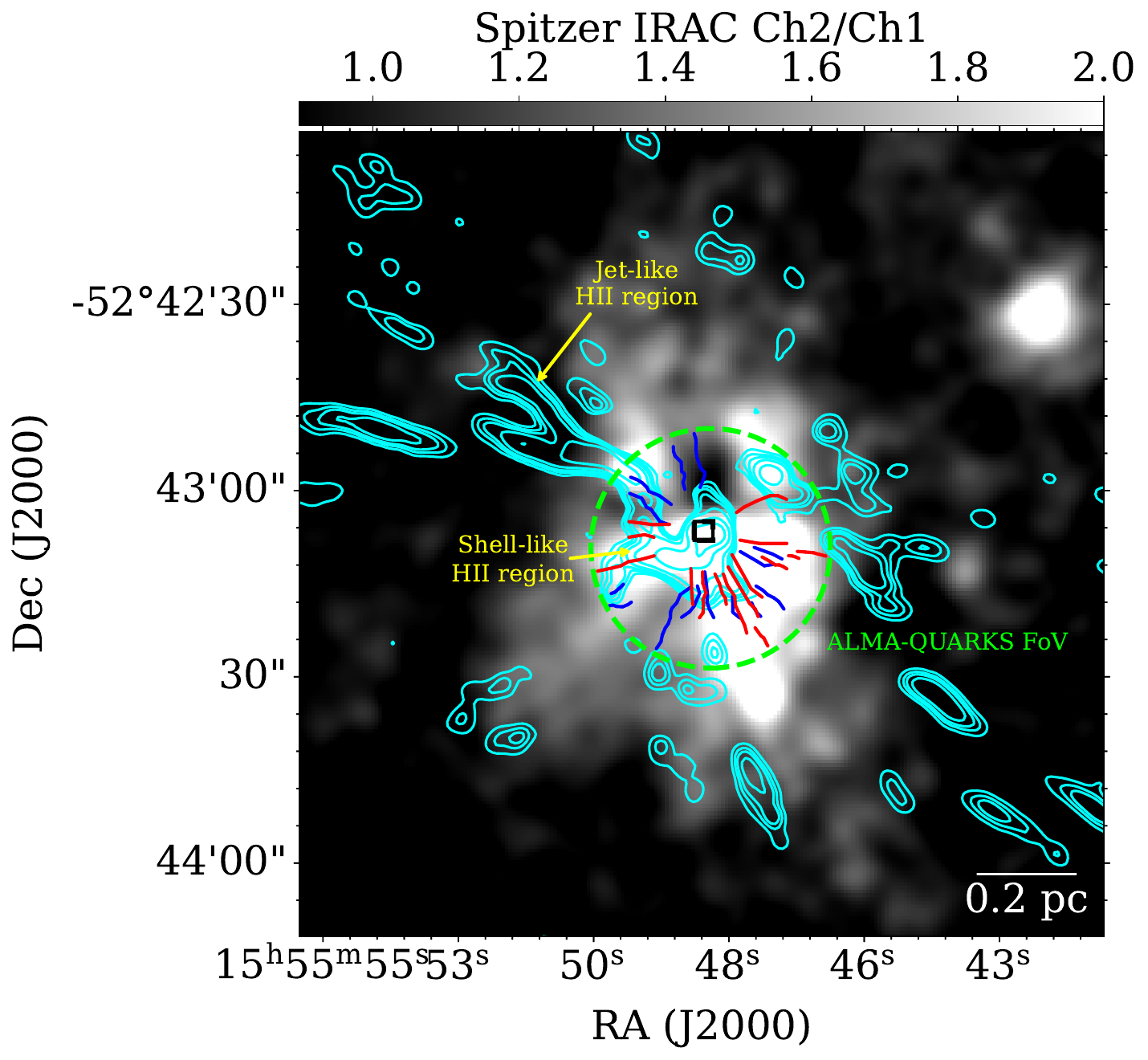}
    \includegraphics[width=0.49\textwidth]{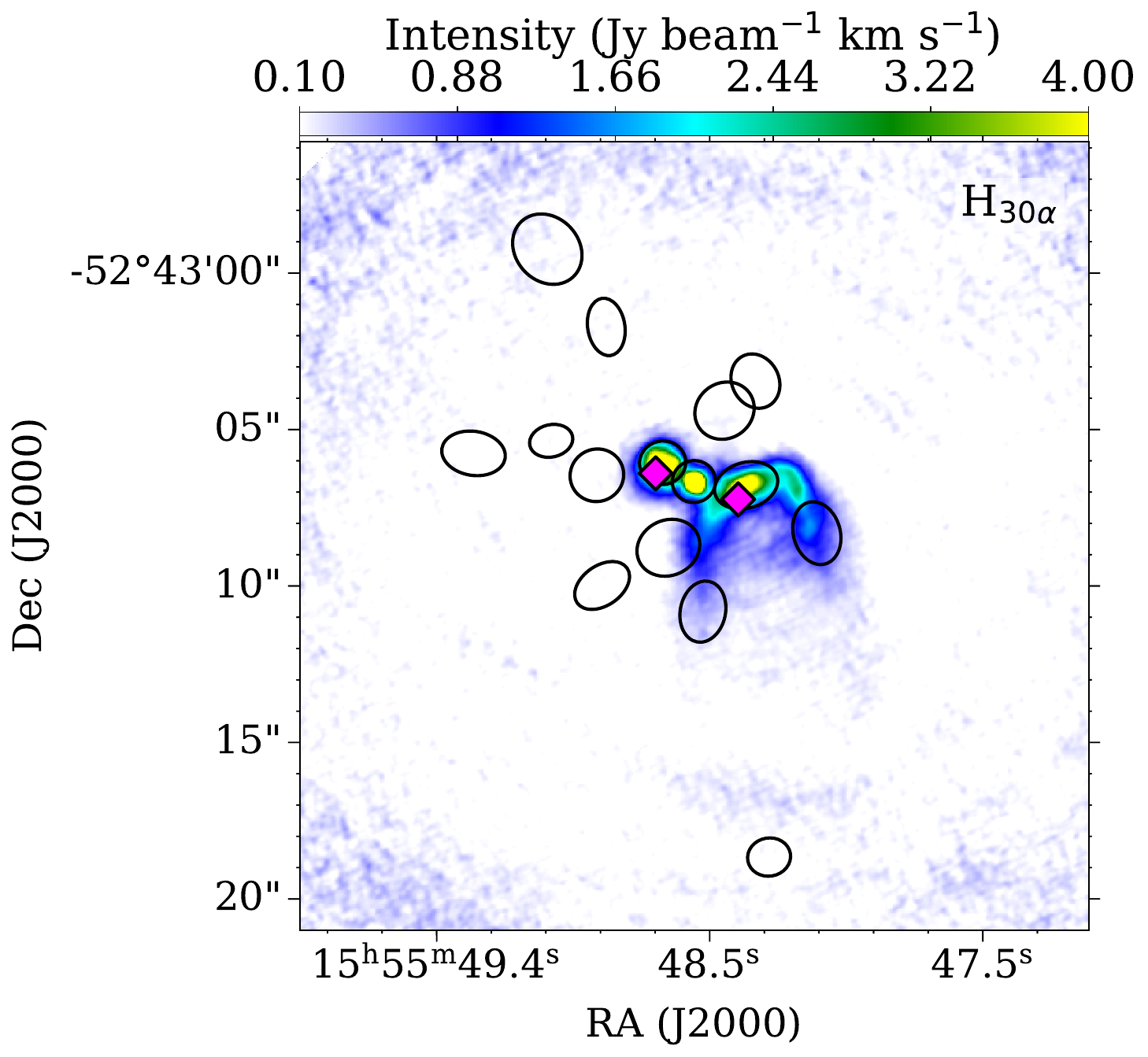}

\caption{(Left) Spitzer ratio Map Ch2(4.5$\mu$m)/Ch1(3.6$\mu$m) of I15520. The cyan contours represent the CORNISH 5.5 GHz radio continuum emission at the levels of [10, 20, 30, 50, 100]$\times\sigma$, where $\sigma=0.2$ mJy/beam. The blue and red lines represent the blue and red-shifted outflow fingers, respectively. The center of explosion is marked with a black rectangle.  The green dashed circle represents the ALMA-QUARKS field of view. A scalebar is shown in the bottom right corner. (Right) H$_{30\alpha}$ emission from the center of the protocluster I15520. The 1.3 mm continuum cores are marked as black ellipses. The magenta diamonds represent the two radio continuum peaks identified in the observations of \citet{Garay2006} using the ATCA.}
\label{fig-ratio_map}
\end{figure*}

We use the {\it Spitzer}-IRAC ratio map Ch2(4.5$\mu$m)/Ch1(3.6$\mu$m) to identify the large-scale shock extent of the explosive source. The IRAC Ch2/Ch1 ratio map is  capable of tracing shocked regions, as the Ch2 of the IRAC band includes shock-excited H$_2$ emission at 4.69 $\mu$m and Br$\alpha$ emissions at 4.05 $\mu$m, while Ch1 contains Polycyclic Aromatic
Hydrocarbon (PAH) features at 3.3 $\mu$m \citep{povich2007}. 
In a detailed analysis of the {\it Spitzer}-IRAC bands toward six Herbig-Haro objects, \citet{takami2010} reported that a flux ratio (Ch2/Ch1) $\geq1.5$ is dominated by shocked emissions, while a flux ratio $\ll1.5$ is indicative of stellar emissions.
The left panel of Figure~\ref{fig-ratio_map} represents the IRAC Ch2/Ch1 ratio map for I15520. The bright extended emission in the ratio map represents the shocked emissions in the region. The cyan contours in Figure~\ref{fig-ratio_map} represent the 5.5 GHz radio continuum emissions associated with the explosive outflow. Two distinct morphologies of the radio continuum emissions can be visible. One toward the explosive center where the radio emission traces a shell of H{\sc ii} region. In the outer regions, away from the center of explosion, the radio emission traces multiple extended jet-like structures. Although the morphology of these extended emission suggests the possibility of candidate ionized jets, further high-resolution and high-sensitivity radio observations are required to confirm the nature of these structures.

In the right panel of Figure~\ref{fig-ratio_map}, we present the H$_{30\alpha}$ emission from the QUARKS survey. The H$_{30\alpha}$ emission traces two hyper compact H{\sc ii} regions and one cometary H{\sc ii} region, concentrated mostly toward the three massive cores located in the center of the cluster. \citet{Garay2006} previously reported the presence of two radio peaks at the center of explosion from the observations at 1.4, 2.5, 4.8, and 8.6 GHz using the Australia Telescope Compact Array (ATCA). In the right panel of Figure~\ref{fig-ratio_map}, we marked the two radio peaks from \citet{Garay2006} with magenta diamonds. The peaks of the radio emission coincide with two of the three H{\sc ii} regions detected in H$_{30\alpha}$ emission. 

\section{Discussion} \label{sec:discussion}
\subsection{Evidence of an Explosive event}\label{sec:discussion explosive}

Our analysis reveals 28 narrow, high-velocity outflow fingers, following the Hubble-Lemaître velocity law, and possibly originating from a single event that occurred at the center of the protocluster. The outflow fingers are distributed quasi-isotropically with spatially overlapped blue- and red-shifted fingers and they do not have any bipolar counterparts. The outflow fingers point back to a common center, which approximately coincides with the center of the protocluster, where most of the massive cores are located.
Regarding the kinematics, some of the blue-shifted and red-shifted fingers are spatially overlapped in projection, one of the key characteristics of explosive outflows. In addition, most outflow fingers show a velocity gradient with increasing velocity far from the center of the protocluster, which is also a distinctive signature of explosive outflows \citep[e.g.,][]{guzman2024, issac2025}. As in these other cases, shorter ejections show larger velocities and vice versa, supporting the tridimensional isotropic configuration of the outflow.

The estimated kinetic energy of the outflow is at least 4.1$\times10^{48}$ erg. Considering a large portion of initial kinetic energy may have been radiated away by the shocks, it suggests that the outflow associated with I15520 is a highly energetic explosive outflow. Its energy is at least one or two orders of magnitude higher than the typical energies of bipolar outflows associated with massive protostars \citep[][and references therein]{zapata2017}. Finally, the mass and momentum of the outflow in I15520 are also similar to those found in other explosive outflows.
Note that, the angular resolution of the data does not allow us to determine whether some of the continuum sources are receding from the common center, as seen in Orion BN/KL. Furthermore, due to the limitations in resolution and sensitivity, we cannot rule out the presence of a few individual bipolar outflows in the region, which are expected in an active star-forming region like, I15520, containing  multiple HMCs. However, the overall kinematics and morphological evidence suggest the explosive nature of the outflow.

\subsection{Expansion of CH$_3$OH Shell at the Center of Explosion}
\begin{figure*}[ht!]
\centering
    \includegraphics[width=0.49\textwidth]{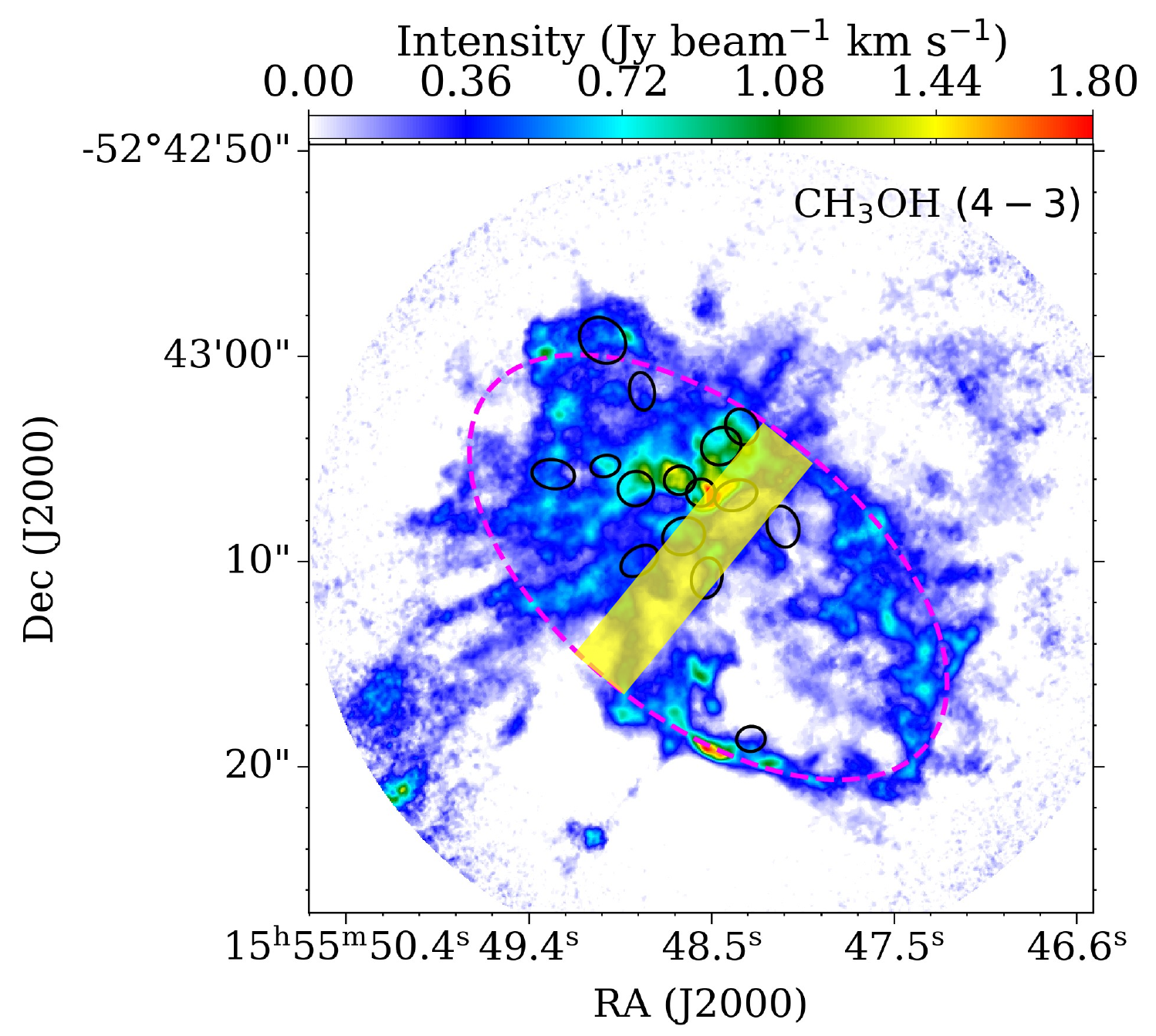}
    \includegraphics[width=0.49\textwidth]{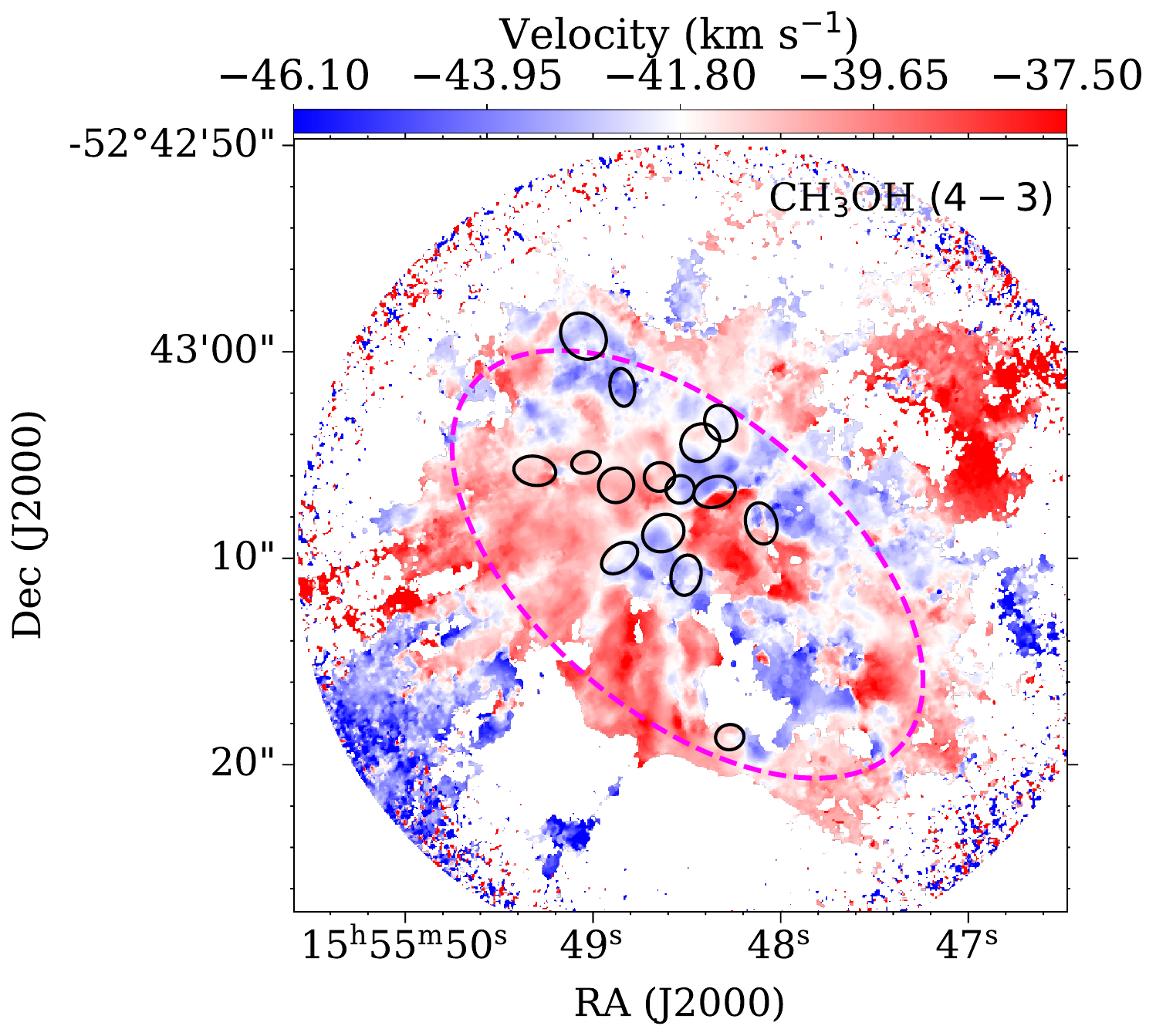}\\
    \includegraphics[width=0.45\textwidth]{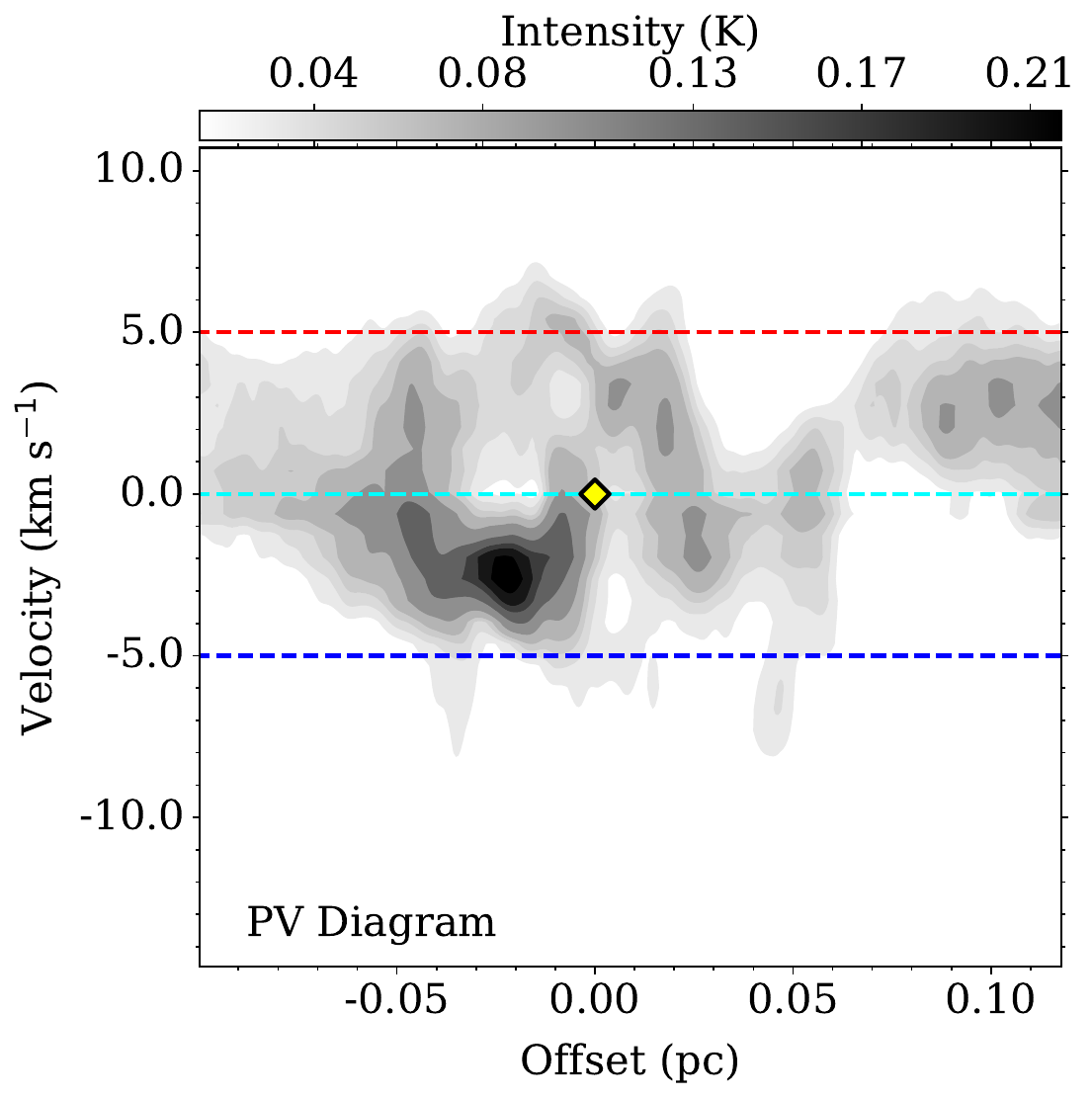}
    \includegraphics[width=0.49\textwidth]{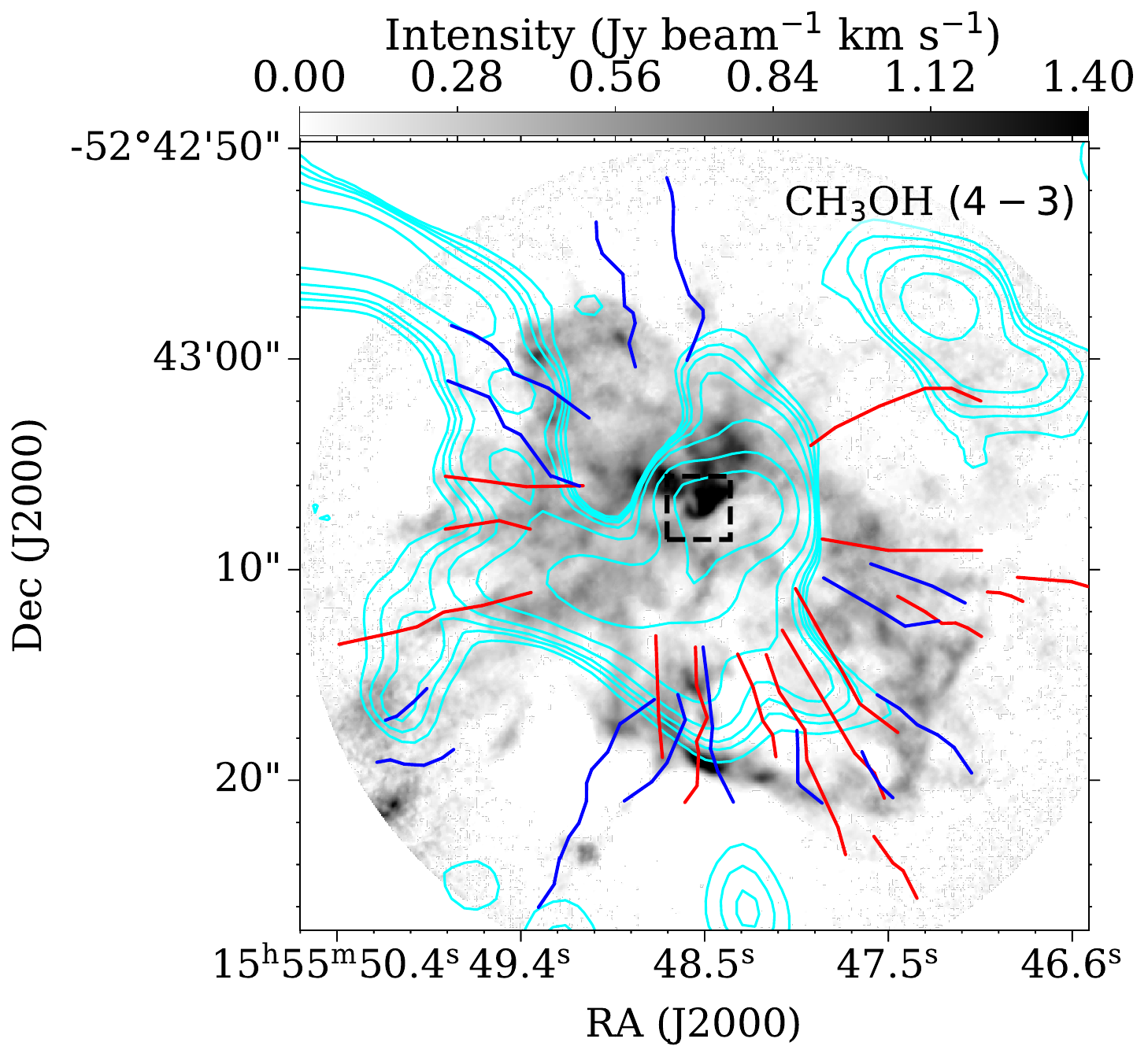}
\caption{(Top) The ALMA CH$_3$OH ($4-3$) moment-0 (left) and moment-1 (right) map of I15520 protocluster. The 1.3 mm continuum cores are shown as black ellipses. The magenta ellipse represents the tentative extent of the expanding CH$_3$OH shell. The yellow rectangle shows the path along which PV diagram is generated. (Bottom left) The PV diagram along the length of the yellow rectangle in the CH$_3$OH moment-0 map generated from north-west to south-east direction. The blue and red dashed lines represent the velocity extent of the blue- and red-shifted expanding cloud with reference to the V$_{\text{LSR}}$, respectively. The cyan dashed line represents the central velocity of the cloud. The center of the expanding shell is marked with a yellow diamond. (Bottom right) CH$_3$OH ($4-3$) moment-0, overlaid with the 5.5 GHz radio continuum contours at the levels of [10, 20, 30, 50, 100]$\times\sigma$, where $\sigma=0.2$ mJy/beam. The blue and red colored lines represent the blue- and red-shifted outflow fingers. The black dashed rectangle represent the center of explosion.}
\label{fig-CH3OH_shell}
\end{figure*}

In dense star-forming regions, CH$_3$OH is known to trace shocked gas, mostly close to the systematic velocity of the cloud \citep{holdship2017, james2020, simone2024}. Therefore, we utilize CH$_3$OH J$=4(2,2)-3(1,2)$ emission (hereafter CH$_3$OH) to investigate its correlation with the explosive event and the shell-like H{\sc ii} region.  In the top left panel of Figure~\ref{fig-CH3OH_shell}, the CH$_3$OH moment-0 map traces a shell-like structure around the explosive center, while the CH$_3$OH moment-1  map (in the top right panel of Figure~\ref{fig-CH3OH_shell}) shows a slightly blue-shifted (toward north-west directions) and red-shifted (toward south-east direction) velocity of the CH$_3$OH shell. A tentative extent of the expanding shell is drawn visually in the top panels of Figure~\ref{fig-CH3OH_shell}. We further investigate the expanding CH$_3$OH shell by generating a PV diagram along a rectangular strip across the shell. Figure~\ref{fig-CH3OH_shell} (Bottom-left) represents the PV diagram across the expanding shell. The PV diagram indicates that the CH$_3$OH shell is slowly expanding outward at velocities $\sim \pm5$ km s$^{-1}$ from the V$_{\text{LSR}}$. However, the center of expansion does not exactly coincide with the center of explosion.

We further investigate the spatial correlation of the CH$_3$OH shell with the 5.5 GHz radio continuum emission and also with the distribution of the outflow fingers. In the bottom right panel of Figure~\ref{fig-CH3OH_shell}, the CH$_3$OH moment-0  map is overlaid with contours (cyan) of  5.5 GHz radio continuum emission and blue- and red-shifted outflow in blue and red colored lines, respectively. We find the distribution of the CH$_3$OH shell correlates well with both the shell of the H{\sc ii} region and also the distribution of the outflow fingers. Therefore, this suggests that the CH$_3$OH shell could have arisen due to compression of the ambient medium driven either by the shell of the H{\sc ii} region, by the explosive outflow, or by the combined effect of both mechanisms.

\subsection{Rate of Explosive Events in the Milky Way}
\begin{figure*}[ht!]
\centering
    \includegraphics[width=0.6\textwidth]{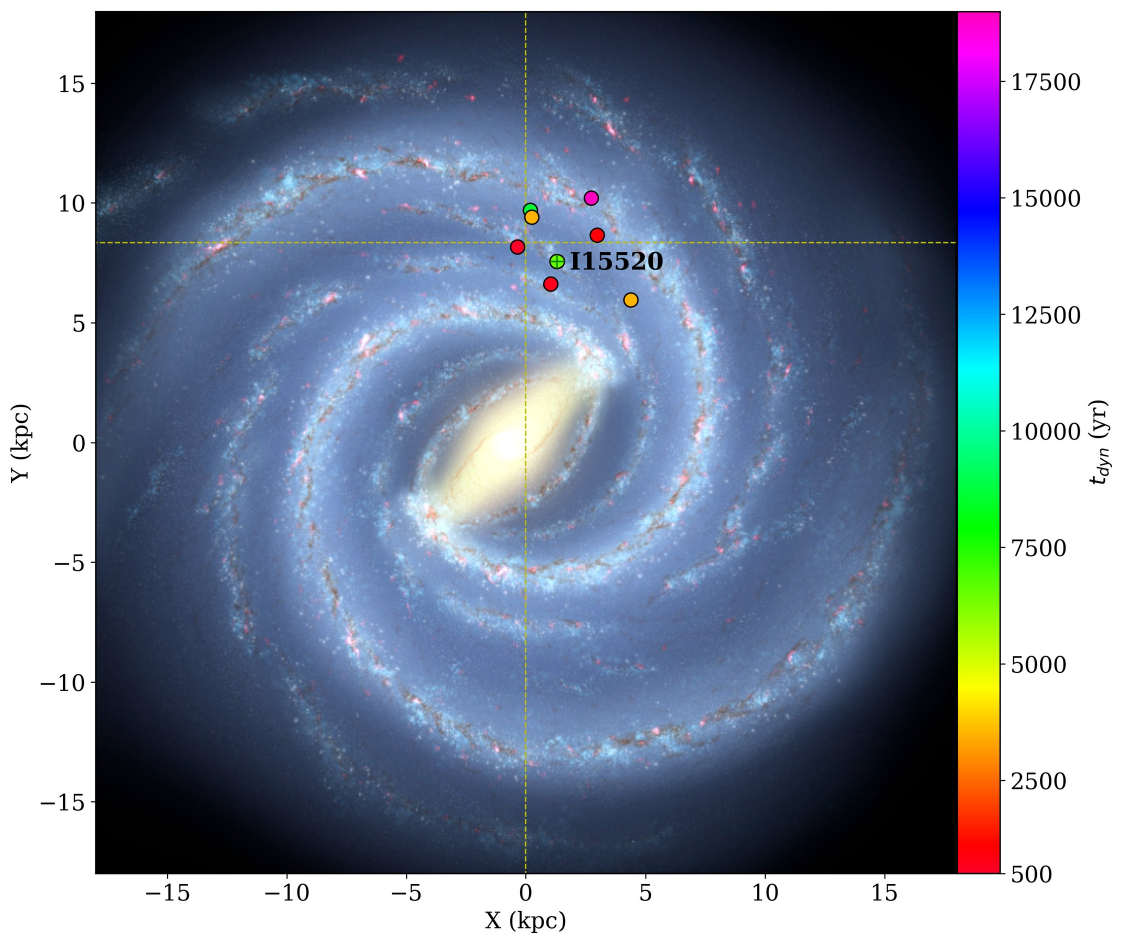}
\caption{Projected spatial distribution of the confirmed explosive outflows in the Milky Way (artist's concept, R. Hurt: NASA/JPLCaltech/Spitzer Science Center). The positions of the explosive outflows are represented by colored dots. I15520 source is marked with a crossed dot. The color of the dots represents the dynamical age of the explosive outflows. The location of the Sun is marked by the crossing of the two dashed lines.}
\label{fig-rateEO}
\end{figure*}

Until now, a total of seven explosive outflows have been reported in the Milky Way Galaxy (Orion BN/KL, DR21, G5.89-0.39, IRAS 16076-5134, Sh2-106, IRAS 12326-6245, and G34.26+0.15). In this study, we have reported a new explosive outflow in I15520. In Figure~\ref{fig-rateEO}, we present the spatial distribution of the confirmed explosive outflows in the Milky Way Galaxy, including our recent addition to the list, I15520. Based on the seven explosive sources identified earlier, \citet[][]{issac2025} estimated the rate of explosive events in the Galaxy to be one every 160 years. However, the rate was earlier estimated by \citet[][]{zapata2023} by considering the first six explosive outflows in the Galaxy as one every 90 years. With the inclusion of the newly detected I15520 explosive outflow, we re-estimate the rate of explosive events following the same method described in \citet[][]{guzman2022,zapata2023}.
We assume that the explosive outflows are evenly distributed in time over a time span of 19000 years covering the time period of all eight explosive outflows (after taking into account their different distances to the Earth) and are distributed within a circle of diameter 5.6 kpc (the separation between the farthest known explosive outflows, IRAS 16076-5134 and DR21). We also assume that the Galaxy is a thin disk with a radius of 15 kpc and estimate the rate of explosive events, which is one event in every 83 years.
Note that the method employed here is simplified and relies on several assumptions, such as the size of the Galaxy, constant star formation rate with galactocentric distance, and a rough estimate of the dynamical timescale of the outflow. The estimate will be more refined with more detections of the explosive outflows in the Galaxy. The estimated rate of explosive outflows is very similar to the massive star formation rate \citep[one star in every 50 years;][]{zapata2023} and the rate of supernovae formation \citep[one in every 50 years;][]{diehl2006, zapata2023}. Therefore, it is possible that massive stars may undergo such explosive phase during their formation and a more comprehensive study would be beneficial to deepen our understanding of massive star formation.
\subsection{Possible Driving Mechanism of Explosive Events}
The driving mechanisms for explosive outflows have been investigated in multiple studies. Among them, the most compelling and observationally supported explanations are the rearrangement of non-hierarchical systems and their dynamical interactions \citep[][]{Bally2011}. A simulation study by \citet[][]{rivera2021} involving the close encounter of a massive star with a cluster of particles also produced outflow fingers that resemble the explosive outflow observed in Orion BN/KL.
In I15520, we find that the continuum cores located within the protocluster are mass segregated, with most of the massive cores located close to the center of the protocluster, and the less massive cores are located in the outer region of the protocluster. The central massive cores (C1, C2, and C3) are separated by a sky-projected distance of $\sim4000$ au, while their sizes ($\sim3500-5400$ au) are also comparable to this distance. Mass segregation in I15520 was also previously reported in \citet[][]{Xu2024} using ALMA band 7 observations. It is possible that the protocluster had undergone a rearrangement of masses, and the dynamical interaction among the central massive cores may result in the formation of explosive outflows. Previous observations of \citet[][]{Garay2006} identified extended H{\sc ii} region at the center of the protocluster. The authors also determined the spectral types of the ionizing sources, and suggested the presence of two massive stars of spectral types O9.5 and B0. Therefore, an encounter of these massive stars with the massive protocluster could also be a possible mechanism for the formation of the explosive event. But the data used in this analysis preclude us from verifying the exact driving mechanism. High-resolution, more sensitive observations are needed to investigate the center of the explosion at scales of $\sim$100-1000 au to confirm the driving mechanism of the explosive event.

\section{Summary} \label{sec:summary}
In this study, we have presented the high-resolution ALMA band 6 observations of the massive star-forming region, I15520. We found 28 well-collimated, molecular outflow fingers, most of which follow a Hubble-Lemaître velocity law. We estimated the total mass, momentum, and energy of the outflow fingers, which are at least 23.8 M$_{\odot}$, 3129.8 M$_{\odot}$ km s$^{-1}$, 4.1$\times10^{48}$ erg, respectively. We found that all the outflow fingers emerged from a common center, which almost coincides with the center of the massive protocluster. The common center also coincides with the position of the UCH{\sc ii} region. These characteristics of the I15520 outflow fingers resemble all the characteristics of explosive outflows previously reported in the literature. We derived the dynamical age of this explosive event $\sim6550$ years. We re-estimated the rate of explosive outflows in the Galaxy, which is one event in $\sim$83 years. We found that the compact cores within the protocluster are mass segregated, with most of the massive cores being at the center. We speculate that the dynamical interactions among these massive cores might be the origin of this explosive event.
\begin{acknowledgments}
We thank the anonymous referee for their valuable comments and suggestions, which greatly improved the quality of the manuscript.
 AH and TB thank the support of the S. N. Bose National Centre for Basic Sciences under the Department of Science and Technology, Govt. of India. AH also thanks the CSIR-HRDG, Govt. of India, for funding the fellowship.
This work was performed in part at the Jet Propulsion Laboratory, which is operated by the California Institute of Technology under contract with the National Aeronautics and Space Administration (80NM0018D0004).
SRD acknowledges support from the Fondecyt Postdoctoral fellowship (project code 3220162) and ANID BASAL project FB210003. NKB acknowledges the support of the China Postdoctoral Science Foundation through grant No. 2025M773187. LB and GG gratefully acknowledges support by the ANID BASAL project FB210003.

This paper utilizes the following ALMA dataset: ADS/JAO.ALMA\#2021.1.00095.S. 
ALMA is a partnership of ESO (representing its member states), NSF (USA), and NINS (Japan), together with NRC (Canada), MOST and ASIAA (Taiwan), and KASI (Republic of Korea), in cooperation with the Republic of Chile. The Joint ALMA Observatory is operated by ESO, AUI/NRAO, and NAOJ.
\end{acknowledgments}

\facilities{ALMA}

\software{astropy \citep{astropy2013}, spectral-cube \citep{spectralcube2019}, APLpy \citep{aplpy2012}, SAOds9 \citep{ds9}}

\bibliography{references}{}
\bibliographystyle{aasjournalv7}

\appendix
%channel maps
\section{CO Channel Maps of the Explosive Outflow in I15520 Region}\label{Appendix-ChannelMaps}
\begin{figure}[ht!]
\figurenum{A1}
\centering
    \includegraphics[width=1.0\textwidth]{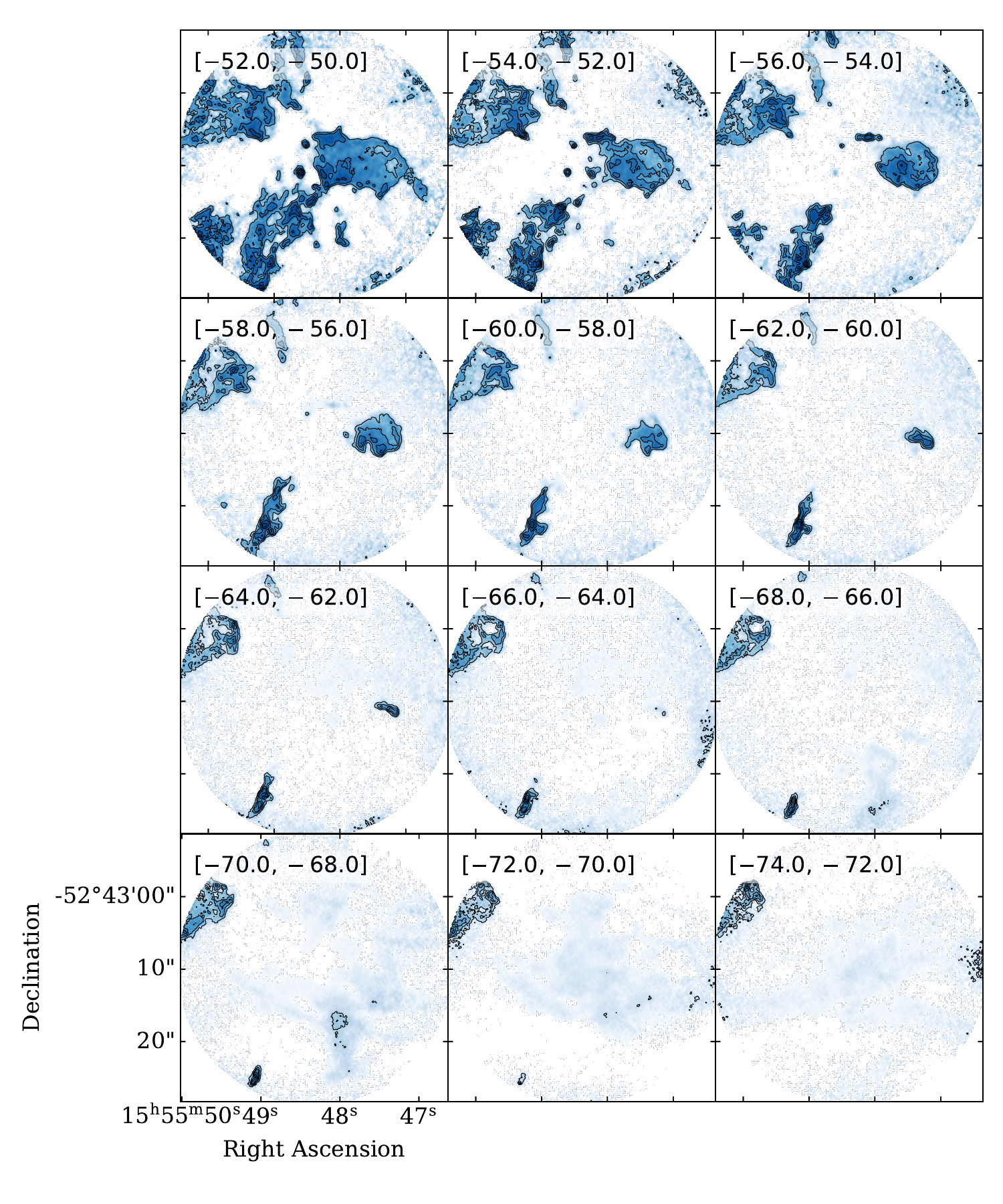}

\caption{The ALMA $^{12}$CO (J$=2-1$) channel maps of the blue-shifted outflow wings in I15520 region. Each panel represents the integrated intensity map within the velocity range as mentioned (in km s$^{-1}$) in each channel.}
\label{Fig:Appendix- blueshifted channel maps}
\end{figure}

\begin{figure}[ht!]
\figurenum{A2}
\centering
    \includegraphics[width=1.0\textwidth]{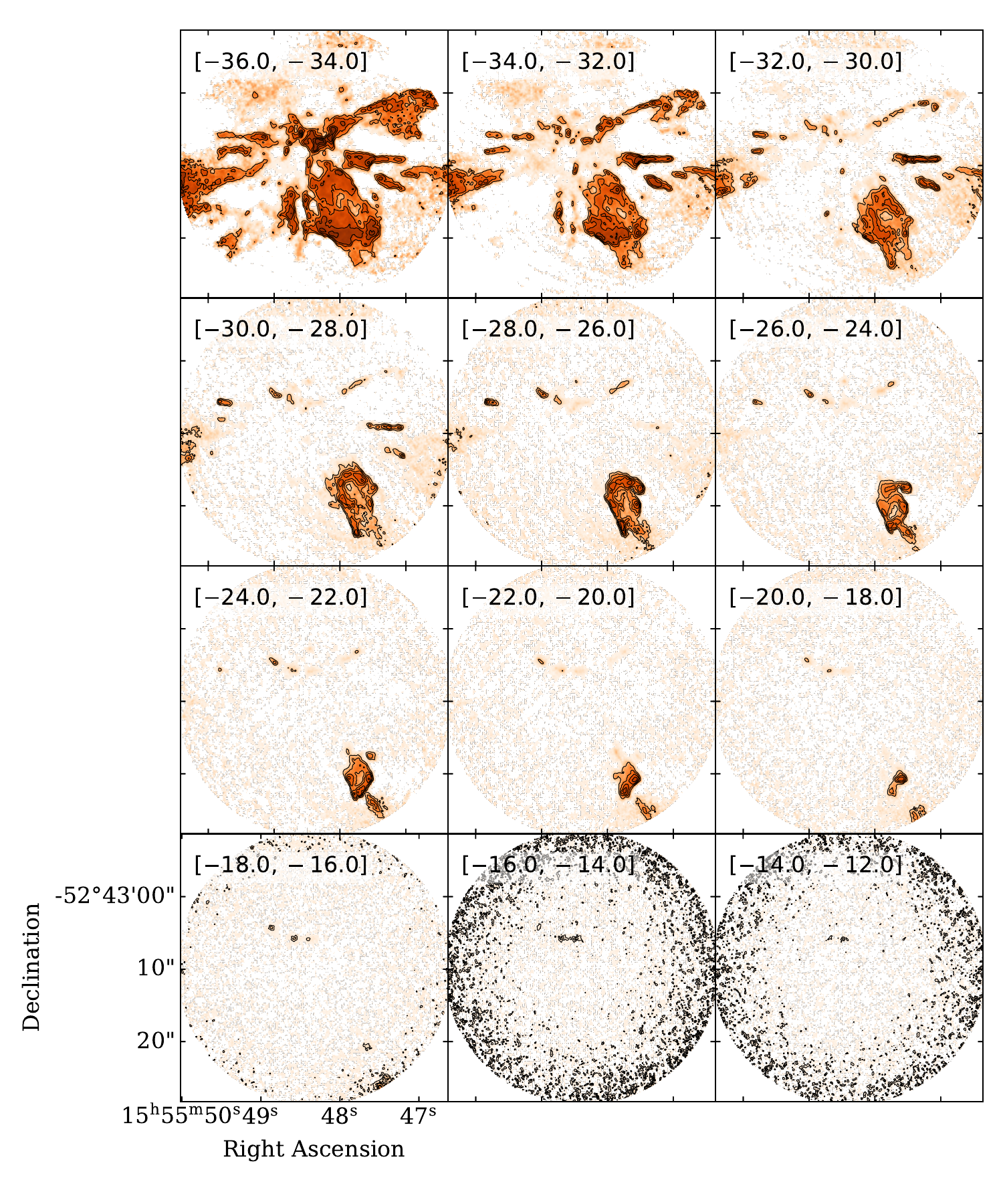}

\caption{The ALMA $^{12}$CO (J$=2-1$) channel maps of the red-shifted outflow wings in I15520 region. Each panel represents the integrated intensity map within the velocity range as mentioned (in km s$^{-1}$) in each channel.}
\label{Fig:Appendix- redshifted channel maps}
\end{figure}

In this section, we present the CO channel maps blue- and red-shifted components of the explosive outflow in the I15520 region in Figure~\ref{Fig:Appendix- blueshifted channel maps} and \ref{Fig:Appendix- redshifted channel maps}. A detailed description of Figure~\ref{Fig:Appendix- blueshifted channel maps} and \ref{Fig:Appendix- redshifted channel maps} is discussed in Section~\ref{subsec:identification}.

\section{Distances of the Compact Cores from the Explosive Center}
\begin{figure}[ht!]
\figurenum{B}
\centering
   \includegraphics[width=0.75\textwidth]{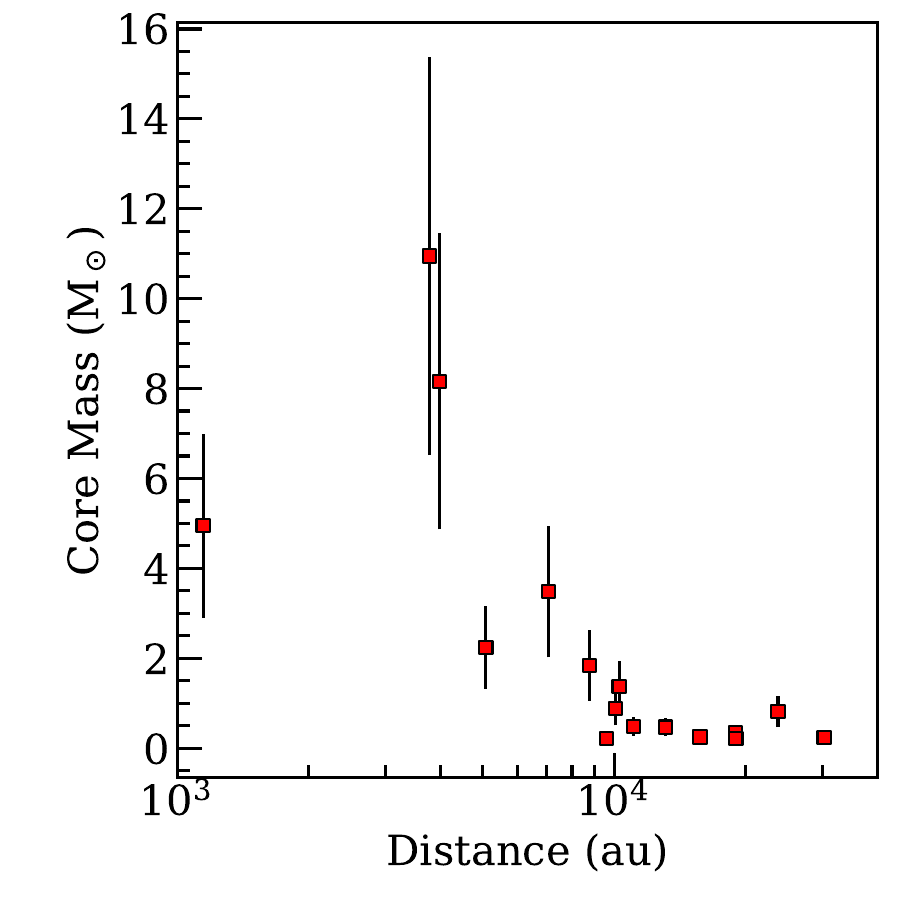}
\caption{Scatter plot of core mass versus projected distance from the explosive center for the compact cores in I15520. The uncertainty in the core mass is shown with black error bars.}
\label{Appendix-MassDistribution}
\end{figure}

In this section, we present the sky-projected distances of the compact cores from the explosive center. Figure~\ref{Appendix-MassDistribution} shows the distribution of core masses with the sky-projected distance from the center of explosion.

\section{Additional Moment Maps of Outflow Tracers}
\begin{figure}[h!]
\figurenum{C}
\centering
   \includegraphics[width=0.95\textwidth]{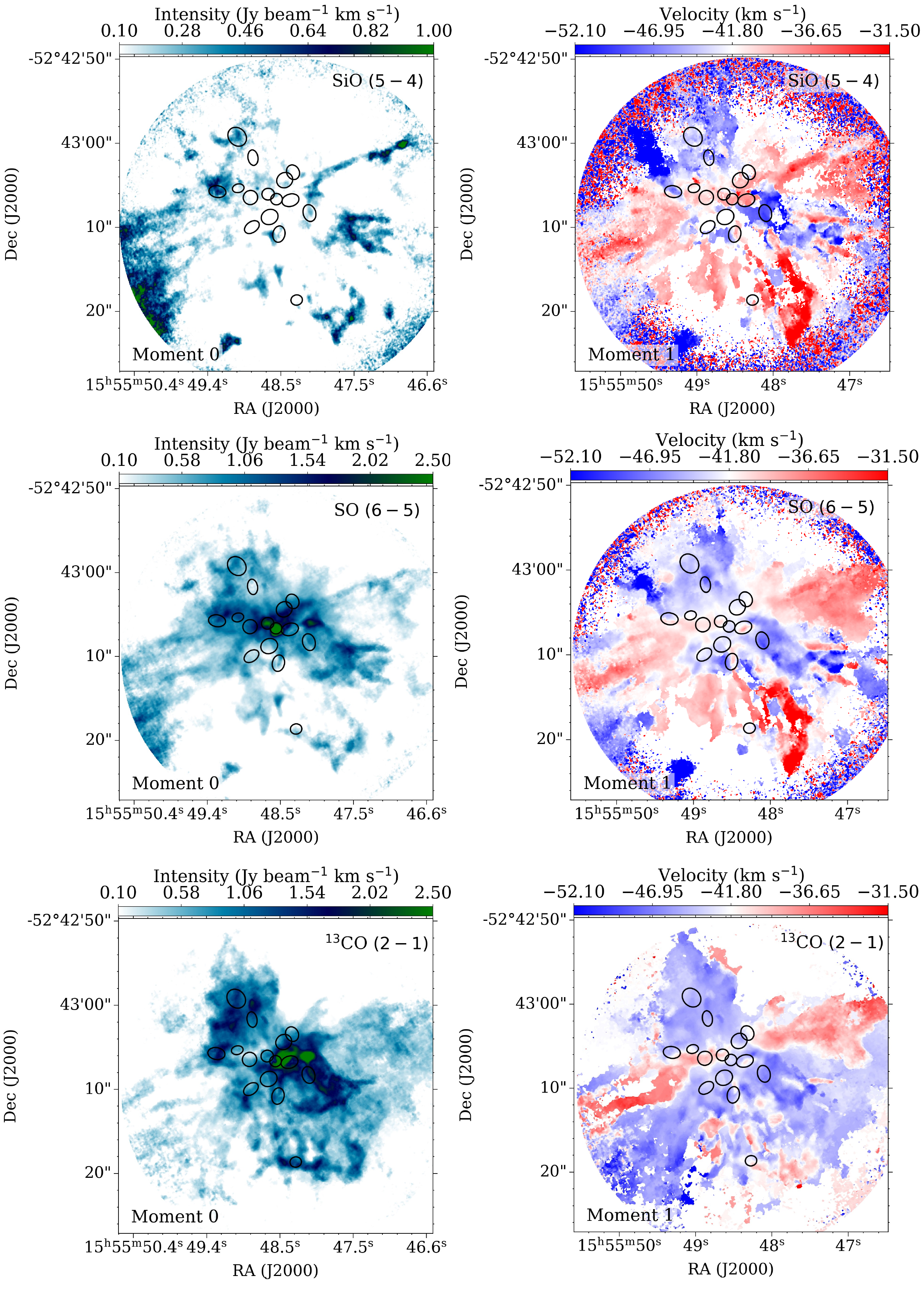}
\caption{Moment 0 (left) and Moment 1 (right) maps of outflow tracers covered in QUARKS survey. The name of the molecular line is quoted in the top left corner of each plot. The 1.3 mm continuum cores are shown as black ellipses.}
\label{Appendix-outflow moment maps}
\end{figure}

In this section, we present the moment 0 and 1 maps of SiO $(5-4)$, SO $(6-5)$, and $^{13}$CO $(2-1)$ molecular lines from the QUARKS survey. In Figure~\ref{Appendix-outflow moment maps} left and right panels, we present the intensity and velocity distribution of the molecular lines, respectively, tracing the morphology and kinematics of the outflow fingers. 

\end{document}